\documentclass[fleqn,usenatbib]{mnras}

\usepackage{newtxtext,newtxmath}

\usepackage[T1]{fontenc}

\DeclareRobustCommand{\VAN}[3]{#2}
\let\VANthebibliography\thebibliography
\def\thebibliography{\DeclareRobustCommand{\VAN}[3]{##3}\VANthebibliography}

\usepackage{graphicx}	
\usepackage{amsmath}	
\DeclareUnicodeCharacter{2212}{\textminus}

\newcommand*    \msun{{\,\rm{M}_{\odot}}}
\newcommand*    \kms{{\,\rm km\,s^{-1}}}
\newcommand*    \pc{{\,\mathrm{pc}}}
\newcommand*    \kpc{{\,\mathrm{kpc}}}
\newcommand*    \myr{{\,\rm Myr}}
\newcommand*    \gyr{{\,\rm Gyr}}
\newcommand*    \mbh{M_{\rm bh}}
\newcommand*    \mhalo{M_{\rm h}}
\newcommand*    \nhalo{N_{\rm h}}
\newcommand*    \mbulge{M_{\rm b}}
\newcommand*    \nbulge{N_{\rm b}}

\newcommand*    \bhratiob{m_{\rm b}/M_{\rm bh}}
\newcommand*    \bhratioh{m_{\rm h}/M_{\rm bh}}
\newcommand*    \tdf{T_{\rm df}}
\newcommand*    \bhsi{R_{\rm infl}}
\newcommand*    \af{a_{\rm f}}
\newcommand*    \ah{a_{\rm h}}
\newcommand*    \taf{T_{\rm f}}

\title[A Multi-Resolution Method for Merging Black Holes]{A multi-resolution method for modelling galaxy and massive black hole mergers}

\author[K. Attard et al.]{
Kate Attard,$^{1}$\thanks{E-mail: k.attard@surrey.ac.uk}
Alessia Gualandris,$^{1}$
Justin I. Read,$^{1}$
Walter Dehnen$^{2,3}$
\\
$^{1}$ School of Mathematics and Physics, Faculty of Engineering and Physical Sciences, University of Surrey, Guildford GU2 7XH, UK \\
$^{2}$ Astronomisches Recheninstitut, Zentrum für Astronomie der Universität Heidelberg, Mönchhofstraße 12-14, D-69120 Heidelberg, Germany \\
$^{3}$ School for Physics and Astronomy, University of Leicester, University Road, LE1 7RH Leicester, UK
}

\date{}

\pubyear{2024}

\begin{document}
\label{firstpage}
\pagerange{\pageref{firstpage}--\pageref{lastpage}}
\maketitle

\begin{abstract}
The coalescence of the most massive black hole (MBH) binaries releases gravitational waves (GWs) within the detectable frequency range of Pulsar Timing Arrays (PTAs) $(10^{-9} - 10^{-6})$ Hz. The incoherent superposition of GWs from MBH mergers, the stochastic Gravitational Wave Background (GWB), can provide unique information on MBH parameters and the large-scale structure of the Universe. The recent evidence for a GWB reported by the PTAs opens an exciting new window onto MBHs and their host galaxies. However, the astrophysical interpretation of the GWB requires accurate estimations of MBH merger timescales for a statistically representative sample of galaxy mergers. This is numerically challenging; a high numerical resolution is required to avoid spurious relaxation and stochastic effects whilst a large number of simulations is needed to sample a cosmologically representative volume. Here, we present a new multi-mass modelling method to increase the central resolution of a galaxy model at a fixed particle number. We follow mergers of galaxies hosting central MBHs with the Fast Multiple Method code \textsc{griffin} at two reference resolutions and with two refinement schemes. We show that both refinement schemes are effective at increasing central resolution, reducing spurious relaxation and stochastic effects. A particle number of $N\geq 10^{6}$ within a radius of $5$ times the sphere of inﬂuence of the MBHs is required to reduce numerical scatter in the binary eccentricity and the coalescence timescale to <30$\%$; a resolution that can only be reached at present with the mass refinement scheme.

\end{abstract}

\begin{keywords}
black hole physics -- galaxies: kinematics and dynamics -- galaxies: nuclei -- galaxies: interactions -- gravitational waves -- methods: numerical
\end{keywords}


\section{Introduction}
The detection of Gravitational Waves (GWs) from two merging black holes by the LIGO scientific collaboration \citep{2016Abbott} has opened a new window on the Universe, providing fresh insights into the astrophysical sources of measurable GWs. The signal has proven the existence of black holes and confirmed the prediction that binaries of black holes coalesce. Since this first detection some 90 events have been observed to date, including both black hole and neutron star mergers \citep{LIGO2021}, providing direct measurements of black hole masses and spins. These have provided new impetus to ongoing efforts to detect GWs from binaries of supermassive black holes (BHBs), which represent the loudest sources of GWs in the Universe \citep{peters1964}.

Supermassive black holes (MBHs; $10^5 \lesssim M \lesssim 10^9)$ are found in the centres of all but the smallest galaxies \citep{FF2005,Kormendy2013}, and their masses correlate tightly with the properties of their host galaxies through scaling relations such as the MBH mass-bulge mass relation and the MBH mass-stellar velocity dispersion relation \citep{Kormendy2013,Reines2015}. MBHs grow over time through gas/star accretion, and via mergers with other MBHs following galactic mergers \citep{Begelman1980}. 
The evolution of BHBs proceeds over three major phases: dynamical friction, hardening and gravitational wave inspiral and coalescence. First, dynamical friction causes MBHs to sink to the centre of the merger remnant until they form a bound pair. The pair then hardens due to three-body interactions with stars during the rapid gravitational slingshot phase \citep{1996Quinlan, 2008Sesana}. Following these strong encounters, stars are ejected with velocities comparable to the binary's orbital velocity and a core is carved in the stellar distribution \citep[e.g.][]{2005Merritt, GualandrisMerritt2012}. This `core scouring' by a BHB is the leading mechanism for explaining the observation of large stellar cores in massive elliptical galaxies \citep{merritt2006}. Once all stars initially on losscone orbits have been ejected and the binary is formally bound and hard (at roughly parsec scale separations for typical $10^8\msun$ MBHs), further hardening relies on efficient repopulation of the losscone through angular momentum diffusion, a mechanism which has been proven efficient in non-spherical models like merger remnants \citep{Vasiliev2014, Vasiliev2015, Gualandris2017, 2021Frigo}. Such `collisionless losscone refilling' is able to drive the BHB to separations (of order milliparsec or less) where emission of GWs commences. Inspiral then proceeds quickly in this final GW phase, leading the MBHs to coalescence.

Mergers of the most massive MBHs ($M>10^7\msun$) are detectable by Pulsar Timing Arrays (PTAs) in the frequency range $(10^{-9} - 10^{-6})$ Hz \citep{Desvignes2016,Reardon2016,2019IPTAPerera}, while MBHs in the mass range $(10^4-10^7)\msun$ will be the main target of the Laser Space Interferometer (LISA) in the frequency range $(10^{-4}-1)$ Hz \citep{Amaro-Seoane2017,Barack2019}. LISA detections will provide accurate estimates of masses, spins and orbital parameters up to a redshift of $z=20$. Unlike electromagnetic radiation, GWs travel essentially unobstructed across space and allow us to probe the early evolution of MBHs and their host galaxies at high redshift \citep{colpi2019}.

Recently, the European Pulsar Timing Array (EPTA),  the North American Nanohertz Observatory for Gravitational Waves (NANOGrav) and the Parks Pulsar Timing Array (PPTA) reported evidence for a spatially correlated stochastic red noise signal consistent with expectations for a stochastic Gravitational Wave Background (GWB) \cite{EPTA2023paperIII, NANOGrav2023, Parkes2023}. This represents the incoherent superposition of unresolved binary mergers, which manifests itself as a long timescale, low frequency common signal across the pulsars in the array.  The astrophysical interpretation of the signal will provide information on the large-scale structure of the Universe, the occupation fraction of MBHs in galaxies, the merger rate of galaxies, scaling relations between MBHs and host galaxies and the efficiency of pairing and merging of BHBs \citep{EPTA2023paperV}.

In this context, it is crucial to model GW sources at low frequencies and estimate detection rates for both current and upcoming instruments. These predictions help constrain expected GW event numbers and aid in the interpretation of any signal and its astrophysical implications \citep{2022AmaroWP}. Current estimates come mainly from semi-analytical models and cosmological simulations, and are affected by significant uncertainties. 

Semi-analytical models have been used to follow the evolution of MBHs and their host galaxies, including dark matter (DM) halos, within the context of the $\Lambda$CDM framework \citep[e.g.][]{Somerville2008,VN2009,Sesana2014,Lacey2016}. Starting from a DM merger tree, structures are evolved using semi-analytic prescriptions. Recent models include prescriptions for MBH growth and feedback, and have been shown to reproduce observed properties such as morphology, colour, star formation rates and luminosity functions at low redshift \citep[e.g.][]{RN2018, Barausse2020}. Being computationally inexpensive relative to $N$-body simulations, they can be used to explore a wide parameter space \citep{2018Lagos,2022IzquierdoVillalba}. However, they rely on simplified assumptions and prescriptions regarding the pairing and evolution of BHBs, with key phases like the binary hardening due to encounters with surrounding stars being modelled based on the results of isolated three-body scattering experiments \citep{2006Sesana,2015Sesana,2022IzquierdoVillalba}. Both low and high-mass BH seeds must be considered for the MBH population at high redshift, as seed masses are currently largely unconstrained \citep{2010Volonteri,2022AmaroWP}. 

By contrast, cosmological $N$-body and hydrodynamic simulations model baryonic and dark matter on large scales from early times, following subsequent galactic mergers and gas accretion simultaneously. Galactic merger mass ratios and encounter geometries are consistently modelled in subsequent events, with realistic dynamical friction times. Massive halos can be seeded with black holes, which then grow over time due to gas accretion and mergers. The inclusion of feedback effects results in gas expulsion, and the quenching of both star formation and MBH growth \citep{DiMatteo2005,Pontzen2017,Ricarte2019,Nelson2019}. Formation of BHBs during mergers is then followed down to the resolution limit, generally set by the choice of the softening length, which is of order $1\kpc$ in Illustris \citep{Genel2014} and of order $300 \pc$ in the recent Illustris TNG50-1 \citep{Dylan2019, Pillepich2019}. In the context of BHB evolution, MBHs have only just begun pairing at these distances, and are still unbound to each other. BHBs are then assumed to merge promptly, without any modelling of the hardening phase. A trade-off between simulation volume and resolution is also present, limiting the statistics of BHBs in higher resolution simulations. 

Predictions for GW detectors have been obtained a posteriori using the properties of merging BHBs in cosmological simulations as input to semi-analytical modelling \citep{Kelley2017, Katz2020, Barausse2020}. These generally include an externally imposed delay to account for the time spent by the binary in the dynamical friction and hardening phase, resulting in significantly fewer mergers, especially for models with high mass seeds, with the lowest LISA rates reaching only a few mergers per $4$ yr nominal mission duration \citep{Barausse2020}. Typical LISA merger rates vary significantly for different MBH assumptions, from $\sim (20-110)\,{\rm yr}^{-1}$ \citep{Sesana2011} to $\lesssim  1\,{\rm yr}^{-1}$ \citep{Katz2020}, but appear higher when MBH triple interactions are taken into account \citep{Bonetti2019}. Current predictions give GWB amplitudes in the PTA frequency range, including characteristic strains of $6.9\times10^{-16}$ from the {\textsc MassiveBlack-II} cosmological simulation \citep{2022Sykes}, of $1.4\times10^{-16}-1.1\times10^{-15}$ from the {\textsc SHARK} semi-analytic model \citep{2022Curylo} and $1.2\times10^{-15}$ from the {\textsc L-Galaxies} model \citep{2022IzquierdoVillalba}.

An alternative approach to ensure higher resolution in the central regions of galaxies where MBHs bind into binaries is provided by high accuracy simulations of individual mergers by means of direct summation simulations \citep[e.g.][]{Gualandris2022}, hybrid collisionless/collisional simulations \citep[e.g.][]{2021Nasim}, and hybrid collisionless/regularised simulations \citep[e.g.][]{2017Rantala}. Direct summation simulations achieve low force errors by brute force calculations of all pairwise gravitational forces, but are limited to particle numbers of order a million by their $\mathcal{O}(N^2)$ computational scaling \citep{GualandrisMerritt2012,Gualandris2022, khan2020}. At these low resolutions, stochastic effects lead to spurious Brownian motion of the BHB \citep{Bortolas2016} as well as a significant scatter in the eccentricity of the binary \citep{Nasim2020}. Increasing resolution while maintaining a bound on force errors is challenging, but has been achieved in the Fast Multiple Method (FMM) code \textsc{griffin} with a sophisticated monitoring of errors and adaptive choice of parameters \citep{Dehnen2014}, which we employ in this work. Here, the MBH-MBH and MBH-star interactions are modelled with direct summation while the star-star interactions are modelled with the more efficient, yet still accurate, FMM. An alternative hybrid approach is implemented in the \textsc{KETJU} code \citep{2017Rantala}. This uses a standard tree code for most force calculations (building upon the \textsc{GADGET-3} collisionless code), but an accurate regularised treatment for the BHB and BHB-star encounters, including relativistic corrections through the Post Newtonian formalism \citep{2019Mannerkoski,2022Mannerkoski}. 

Modelling galaxies and galactic mergers at the large resolutions $N\gtrsim 10^7$ required to avoid stochastic effects \citep{Nasim2020} is challenging even with more efficient codes like \textsc{griffin} due to the large range of spatial and temporal scales involved. In this paper, we present a numerical scheme to increase resolution in the central regions of galaxies based on mass refinement. We present results of both isolated galaxy models and merging galaxies and show that mass refinement schemes are effective at increasing resolution in galactic centres, without any adverse effects on stability or dynamical evolution. In particular, we show that a mass refined $N=10^6$ galaxy model behaves similarly to a $N=10^7$ reference galaxy model.

This paper is organised as follows: we first describe the multi-mass resolution scheme in Sec.~\ref{sec:scheme}, the methodology and simulations used in Sec.~\ref{sec:Methods}, and present the results of the mass refined models in Sec.~\ref{sec:evolution}. In Sec.~\ref{sec:orbev} we extrapolate the evolution to late times and estimate the total coalescence timescales. We conclude with a discussion and summary in Sec.~\ref{sec:results}.

\section{A multi-mass resolution scheme}
\label{sec:scheme}
We implement a multi-mass refinement scheme to increase the resolution in the central regions of galaxy models, with the aim to accurately model the evolution of BHBs formed in galactic mergers at a reduced computational cost. In the case of multi-component models featuring a central MBH, a stellar bulge, and a DM halo, the scheme is applied independently to the bulge and the halo.

Starting from a desired total particle number for each component, which we denote $\nbulge$ and $\nhalo$, stars are first oversampled by a factor $k$, termed the multiplication factor, and then coarse-grained based on radial position. A number of radial zones can be defined such that stars in the innermost zone are left unchanged, while only a given fraction of stars is retained in the outer zones. This fraction decreases moving away from the galaxy centre. The mass of each particle in the outer zones is then increased by the same factor $k$ to ensure that the total mass in the galaxy and the mass density profile are unaffected. We adopt a larger number of zones for higher multiplication factors in order to reduce the increase in particle masses at the outermost radii. In this way, we increase the resolution in the central region of the galaxy whilst avoiding large jumps in particle masses between the radial zones, which would otherwise induce spurious mass segregation. A commonly adopted mass factor to avoid mass segregation is of order $10$ \citep{hopman2009}, and this is consistent with the results shown in Fig.~\ref{fig:tdynI}.

We implement two different schemes, labelled `a' and `b', with parameters as listed in Table \ref{tab:schemeparams}. Scheme `a' is our standard implementation, characterised by $n=3$ zones, while scheme `b' is a more aggressively refined scheme with $n=4$ zones. The latter scheme can be considered physically motivated, as its innermost radial bin is equal to twice the radius of the sphere of influence of the central MBH, one of the fundamental scales in the system \citep{Peebles72}
\begin{equation}
    \bhsi = GM_{\rm bh} / \sigma^{2} \,,
	\label{eq:rbhinf}
\end{equation}
where $M_{\rm bh}$ represents the MBH mass and $\sigma$ the local 1D stellar velocity dispersion. 
\begin{table}
	\centering
	\caption{The parameters of the refinement scheme: scheme identifier, zone number $z$ (for both bulge and halo), percentage of total particle number within zone $p$,  mass multiplication factor $m$, and start and end radius ($r_{0}$-$r_{1}$, $\kpc$) of each zone, for bulge and halo.}
	\label{tab:schemeparams}
	\begin{tabular}{llcccccc} 
		\hline
		Scheme & $z$ & $p$ & $m$ & $r_{b0}$ & $r_{b1}$ & $r_{h0}$ & $r_{h1}$ \\
		\hline
		a & 1 & 50.0 & 1 & 0.0 & 1.06 & 0.0 & 18.70 \\
		 & 2 & 37.5 & 5 & 1.06 & 7.15 & 18.70 & 62.80 \\
		 & 3 & 12.5 & 25 & 7.15 & $-$ & 62.50 & $-$ \\
        \hline
		b & 1 & 1.0 & 1 & 0.0 & 0.18 & 0.0 & 4.98 \\
		 & 2 & 49.0 & 2.53 & 0.18 & 1.31 & 4.98 & 21.50 \\
		 & 3 & 37.5 & 10 & 1.31 & 7.15 & 21.50 & 62.80 \\
		 & 4 & 12.5 & 40 & 7.15 & $-$ & 62.80 & $-$ \\
	\hline
	\end{tabular}
\end{table}
The radial boundaries and mass multiplication factors are given in Table~\ref{tab:schemeparams}.  
In scheme ‘a’, particles are initially over-seeded by a factor 5, with the first (innermost) zone retaining all particles, and its radius set by the requirement of containing $50\%$ of all particles ($N_b$ and $N_h$, respectively, for bulge and halo). The particle number is then reduced by a factor of $5$ for each successive radial zone, and the multiplication factor for the remaining particle masses increases correspondingly. This multiplication factor determines the placement of the boundary between the outer radial zones. Scheme `b' is similar, with an additional zone being added to prevent an excessively large increase in mass in the outermost radial zone. In this scheme, the innermost $50\%$ of the particles is divided between zone 1 (with $1\%$) and zone 2 (with $49\%$), while the outermost $50\%$ is divided between zone 3 (with $37.5\%$) and zone 2 (with $12.5\%$).

\section{Methods}
\label{sec:Methods}
\subsection{Galaxy models}
We consider a set of multi-component models representative of massive elliptical galaxies consisting of a stellar bulge, a dark matter (DM) halo, and a central MBH. The stellar bulge follows a Sérsic profile \citep{sersic1963,sersic1968} in projection
\begin{equation}
    I = I_{0}\,e^{-b_{n}(R/R_{e})^{1/n}}
	\label{eq:sersic}
\end{equation}
where $I_{0}$ is the surface intensity, $R_{e}$ is the scale radius, $n$ is the Sérsic index, and $b_{n}$ is a function of $n$, where $b_{n} \approx 2n - \frac{1}{3} + \frac{4}{405n}$ \citep{Ciotti1999}. We set $n = 4$, appropriate for elliptical galaxies, as in the de Vaucouleurs profile \citep{1948dV}.
The DM halo follows an NFW profile \citep{NFW1996}
\begin{equation}
    \rho(r) = \frac{\rho_{0}}{(r/a)(1+r/a)^{2}} 
	\label{eq:nfw}
\end{equation}
where $a$ is the scale radius and $\rho_{0}$ is a normalisation factor. 

We extract merger trees from the IllustrisTNG-300-1 cosmological simulation \citep{Pillepich2019,2018Springel,2018Marinacci,Nelson2019,2018Naiman} and select a major merger at low redshift whose central MBH is an expected PTA source. The MBH has a mass $\mbh=7.14\times 10^{8}\msun$. We then set the parameters for the bulge and halo components using observational scaling relations. The stellar bulge mass $\mbulge$ is derived from the MBH-stellar mass relation
\begin{equation}
    {\rm log}\left(\frac{\mbh}{\mbulge}\right)=\alpha+\beta \,{\rm log} \left(\frac{\mbulge}{10^{11}\msun}\right)
	\label{eq:MBH-Mstellar}
\end{equation}
 \citep{Kormendy2013,Reines2015} with parameters  $\alpha = 8.95 \pm 0.09$ and $\beta= 1.40 \pm 0.21$ taken from fits to observed elliptical galaxies \citep{Reines2015}. This gives $\mbulge=8.54\times10^{10}\msun$.
 
 The halo mass $\mhalo$ is taken from the halo-stellar mass scaling relation of \citet{2014Chae} for massive elliptical galaxies, with a typical value $\mhalo=5.04\times10^{12}\msun$. The virial radius $R_{200}$ is derived from the relation with the halo mass $\mhalo=200\rho_{c}\frac{4}{3}\pi R_{200}^{3}$, with $\rho_{c} = 136.05\msun \kpc^{-3}$ \citep{2008BT}, which gives $R_{200}=353.69\kpc$. The scale radius of the NFW profile is given by $a = R_{200}/c$, where the concentration parameter $c$ is derived from the $c-\mhalo$ relation in \citet{2014Dutton}. Lastly, we set the half-mass radius of the stellar component using the relation $R_{1/2}=0.015R_{200}$ \citep{Kravtsov2013}, giving $R_{1/2}=5.30\kpc$. 

We utilise \textsc{Agama}, an action-based galaxy modelling software library \citep{Vasiliev2019}, to sample the total gravitational potential of the multi-component galaxy system and construct equilibrium initial conditions for our simulations. We adopt units such that $G=1$, $R=1\kpc$ and $M=\mbulge$. We generate isolated models, denoted with identifier `I', at two reference resolutions, $N=10^6$ and $N=10^7$, and apply the mass refinement schemes described in section \ref{sec:scheme} for each resolution. We then set up equal mass mergers of two independent realisations of the same multi-component model, denoted with identifier `M', at the same resolutions. Galaxies are placed on bound Keplerian orbits of eccentricity $e=0.7$ and at an initial distance of $D=378.6\kpc$. The initial eccentricity is relatively high, in agreement with typical orbits in large scale cosmological simulations \citep{KB2006}, but not so high to result in a head-on galaxy collision. The initial distance is set to $R \simeq 9\,R_{s}$, where $R_s$ represents the scale radius of the NFW model describing the halo of the galaxies, defined as the radius where the particle number drops to half of the total. 
The parameters of all models are listed in Table \ref{tab:initconds}, including the total particle number $N$, the number of halo and bulge particles, and the ratio of halo/bulge particle mass to the MBH mass. The final column gives the numerical multiplier $k$ adopted in the mass refined schemes. 

 \begin{table*}
	\centering
	\caption{Properties of the adopted galaxy models: model identifier, type of model (isolated or merger), total particle number, number of halo particles, number of bulge particles, number of particles within the $5\bhsi$ in the initial models, minimum halo star to MBH mass ratio, minimum bulge star to MBH mass ratio, and central resolution multiplication factor for the models with mass refinement.} 
	\label{tab:initconds}
	\begin{tabular}{lcccccccc} 
		\hline
		Run & Type & N & $\nhalo$ & $\nbulge$ & $N(5\bhsi)$ & $\bhratioh$ & $\bhratiob$ & Scheme\\
		\hline
		I6 & Isolated & $10^{6}$ & $9\times10^{5}$ & $1.05\times10^{5}$ & $1.3\times10^{4}$ & $7.84\times10^{-3}$ & $1.14\times10^{-3}$ & -\\
		I7 & Isolated & $10^{7}$ & $9\times10^{6}$ & $1.05\times10^{6}$ & $1.3\times10^{5}$ & $7.84\times10^{-4}$ & $1.14\times10^{-4}$ & -\\
		I6a & Isolated & $10^{6}$ & $9\times10^{5}$ & $1.05\times10^{5}$ & $5.7\times10^{4}$ & $1.57\times10^{-3}$ & $2.38\times10^{-4}$ & Multi-mass ($k=5$)\\
        I6b & Isolated & $10^{6}$ & $9\times10^{5}$ & $1.05\times10^{5}$ & $6.0\times10^{4}$ & $7.84\times10^{-4}$ & $1.14\times10^{-4}$ & Multi-mass ($k=10$)\\
		I7a & Isolated & $10^{7}$ & $9\times10^{6}$ & $1.05\times10^{6}$ & $5.7\times10^{5}$ & $1.57\times10^{-4}$ & $2.27\times10^{-5}$ & Multi-mass ($k=5$)\\
        I7b & Isolated & $10^{7}$ & $9\times10^{6}$ & $1.05\times10^{6}$ & $5.9\times10^{5}$ & $7.84\times10^{-5}$ & $1.14\times10^{-5}$ & Multi-mass ($k=10$)\\
		M6 & Merger & $10^{6}$ & $9\times10^{5}$ & $1.05\times10^{5}$ & $2.5\times10^{4}$ & $7.84\times10^{-3}$ & $1.14\times10^{-3}$ & - \\
		M7 & Merger & $10^{7}$ & $9\times10^{6}$ & $1.05\times10^{6}$ & $2.5\times10^{5}$ & $7.84\times10^{-4}$ & $1.14\times10^{-4}$ & - \\
		M6a & Merger & $10^{6}$ & $9\times10^{5}$ & $1.05\times10^{5}$ & $1.1\times10^{5}$ & $1.57\times10^{-3}$ & $2.38\times10^{-4}$ & Multi-mass ($k=5$)\\
		M6b & Merger & $10^{6}$ & $9\times10^{5}$ & $1.05\times10^{5}$ & $1.2\times10^{5}$ & $7.84\times10^{-4}$ & $1.14\times10^{-4}$ & Multi-mass ($k=10$)\\
		M7a & Merger & $10^{7}$ & $9\times10^{6}$ & $1.05\times10^{6}$ & $1.1\times10^{6}$ & $1.57\times10^{-4}$ & $2.27\times10^{-5}$ & Multi-mass ($k=5$)\\
		M7b & Merger & $10^{7}$ & $9\times10^{6}$ & $1.05\times10^{6}$ & $1.2\times10^{6}$ & $7.84\times10^{-5}$ & $1.14\times10^{-5}$ & Multi-mass ($k=10$)\\
		\hline
	\end{tabular}
\end{table*}

While the ratio between a halo particle mass and a bulge particle mass is equal for all models ($m_h/m_b \sim 6.7$), the MBH to bulge particle mass ratio at the innermost radial shell decreases as the refinement scheme becomes more aggressive. In particular, I6b has the same $\bhratiob$ as I7, and the smallest ratio is obtained for I7b. For all the merger models with $N=10^{6}$, we produce 7 different random realisations, denoted {\it i}  to {\it vii}, in order to measure the scatter in the orbital parameters.

\subsection{Numerical simulations}
We evolve all models with \textsc{griffin} \citep{Dehnen2014}, a Fast Multiple Method (FMM) code that monitors force errors and utilises adaptive parameters to ensure a distribution of force errors similar to that in a direct summation code, while retaining the $\mathcal{O}(N)$) scaling of the FMM technique. The code has been shown to perform similarly to a direct summation code in modelling the evolution of BHBs formed in galactic mergers \citep{Nasim2020}. 

The simulations reported here use a multipole expansion order $p = 5$ and a relative total force error of $\lesssim 1\times10^{-3}$ for the $N=10^{6}$ models and $\lesssim 5\times10^{-5}$ for the $N=10^{7}$ models. The softening parameter is initially set to $\epsilon=30\pc$ for star-star interactions and to $\epsilon_{\rm bh} = 10\pc$ for MBH-MBH and MBH-star interactions for all models. We then reduce the softening length $\epsilon_{\rm bh}$ in the merger simulations to $5\pc$ after a time $\sim 2\gyr$ corresponding to just prior to the end of the dynamical friction phase. This setup allows to reduce computational time in the early stages of the merger, driven by dynamical friction, and yet to accurately model the evolution of the BHB through the rapid hardening phase dominated by encounters with background stars. The softening kernel is a near-Plummer variation for smooth source particles with a density $\rho \propto (r^{2}+\epsilon^{2})^{-3.5}$. All simulations are evolved until the binary reaches a separation of order the softening length $\epsilon_{\rm bh}$, below which the dynamical evolution is no longer reliable. We ensure that this separation is always smaller than the hard-binary separation. 

\section{Evolution of the multi-mass models}
\label{sec:evolution}
Mass refinement schemes of the type presented in this work have been adopted in the literature with the aim of increasing resolution in the central regions of a single galaxy model, with \citet{Cole2012} allowing a reduction in the total particle number by half at the same central resolution without any adverse effects. Here we test the viability of our multi-zone multi-mass schemes, for both isolated and merger models. Galactic mergers are violent events that affect the distribution of stars at both large and small scales, and as such it is not guaranteed that the refinement will work over time in a merger simulation. However, the `mixing theorem' shows that $N$-body simulations should preserve central phase space density, even through mergers \citep{Dehnen2005}. As such, we can anticipate that the high resolution regions at the centre of each merging galaxy should overlap once the merger is complete.

\begin{figure}
\centering
	\includegraphics[width=\columnwidth]{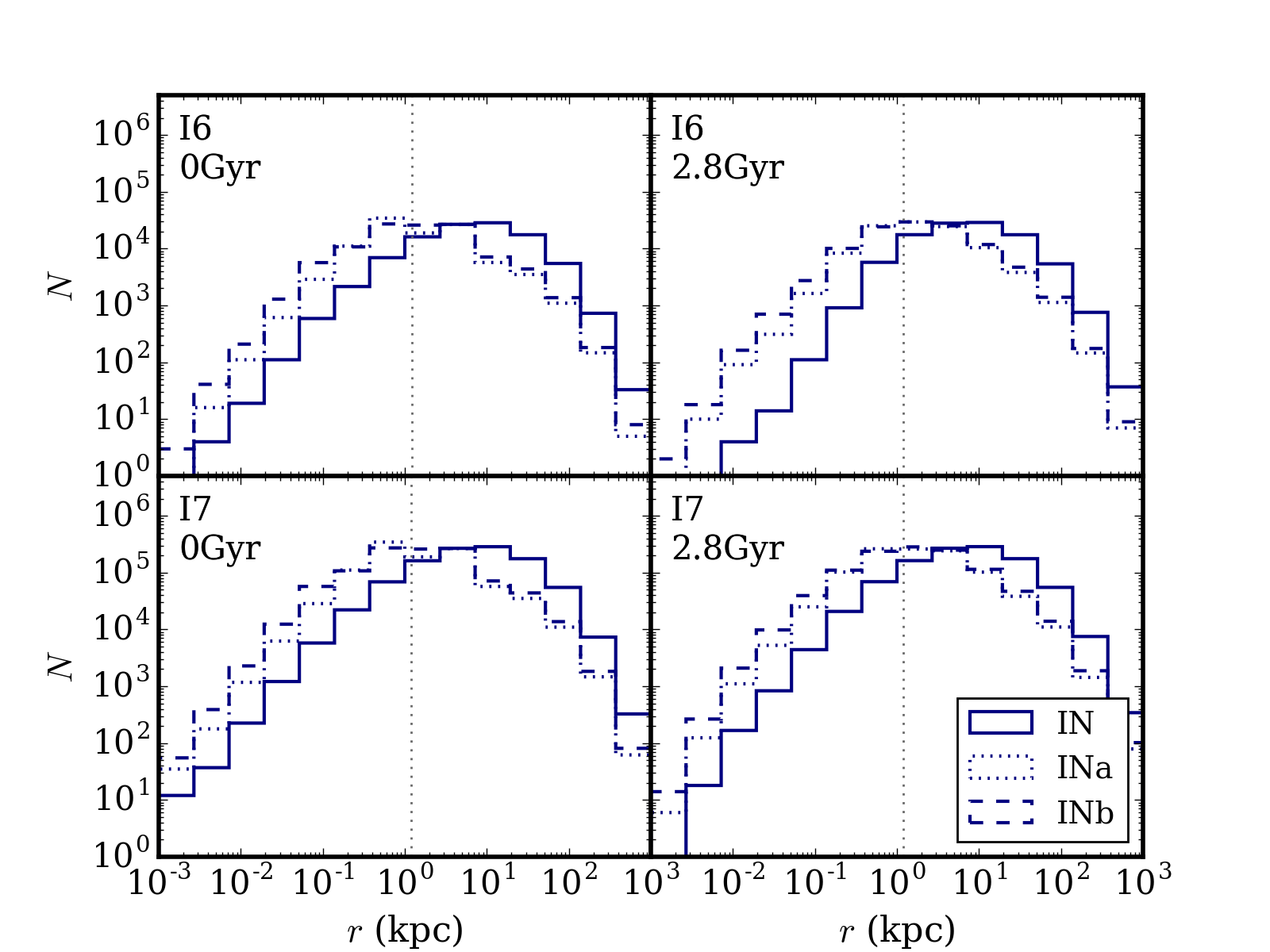}
	\caption{Radial distribution of the bulge particles for all isolated models with $N=10^{6}$ particles (top panels) and $N=10^{7}$ (bottom panels) at two characteristic times: $t=0\gyr$ (left) and $t=2.8\gyr$ (right), corresponding to a late time in the evolution. The grey dotted line represents the $5\bhsi$ radius. A significant increase in the particle number $<10\kpc$ is observed for the models with the multi-mass scheme applied, and no significant expansion occurs over time for these models.}
    \label{fig:INhists}
\end{figure}

\begin{figure}
\centering
	\includegraphics[width=\columnwidth]{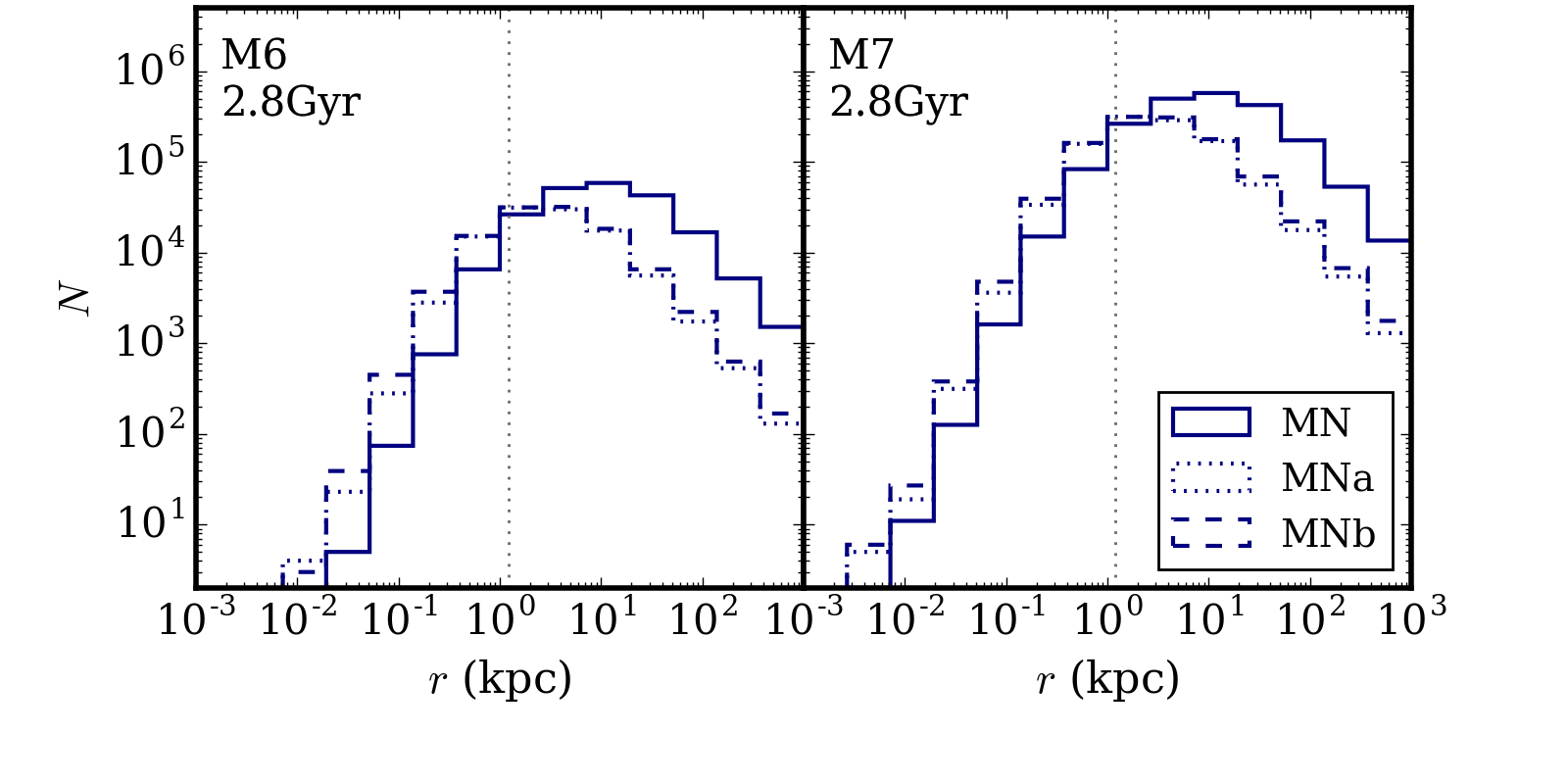}
	\caption{Radial distribution of the bulge particles for all merger models with $N=10^{6}$ particles (left) and $N=10^{7}$ (right), both at the later time $t=2.8\gyr$ past the end of the merger process.  The grey dotted line represents the $5\bhsi$ radius. A higher central resolution is maintained for the models with the multi-mass scheme applied post-merger, most significantly in the M6 models.}
    \label{fig:MNhists}
\end{figure}

The distribution of bulge particles in the isolated models, with and without mass refinement, is shown in Fig.~\ref{fig:INhists} at an early time and a late time in the evolution. This illustrates the efficiency of the refinement schemes, with increased population in the central regions of models 'a' and 'b'. The mass refined models retain their higher central resolution to later times, in agreement with earlier studies \citep{Cole2012}. Crucially, this holds true in the merger models as well, as shown in Fig.~\ref{fig:MNhists}, despite the violent mixing happening during the merger process. Models M6a and M6b achieve a population within the central $\sim100\pc$ similar to model I7. 


\begin{figure}
\centering
	\includegraphics[width=\columnwidth]{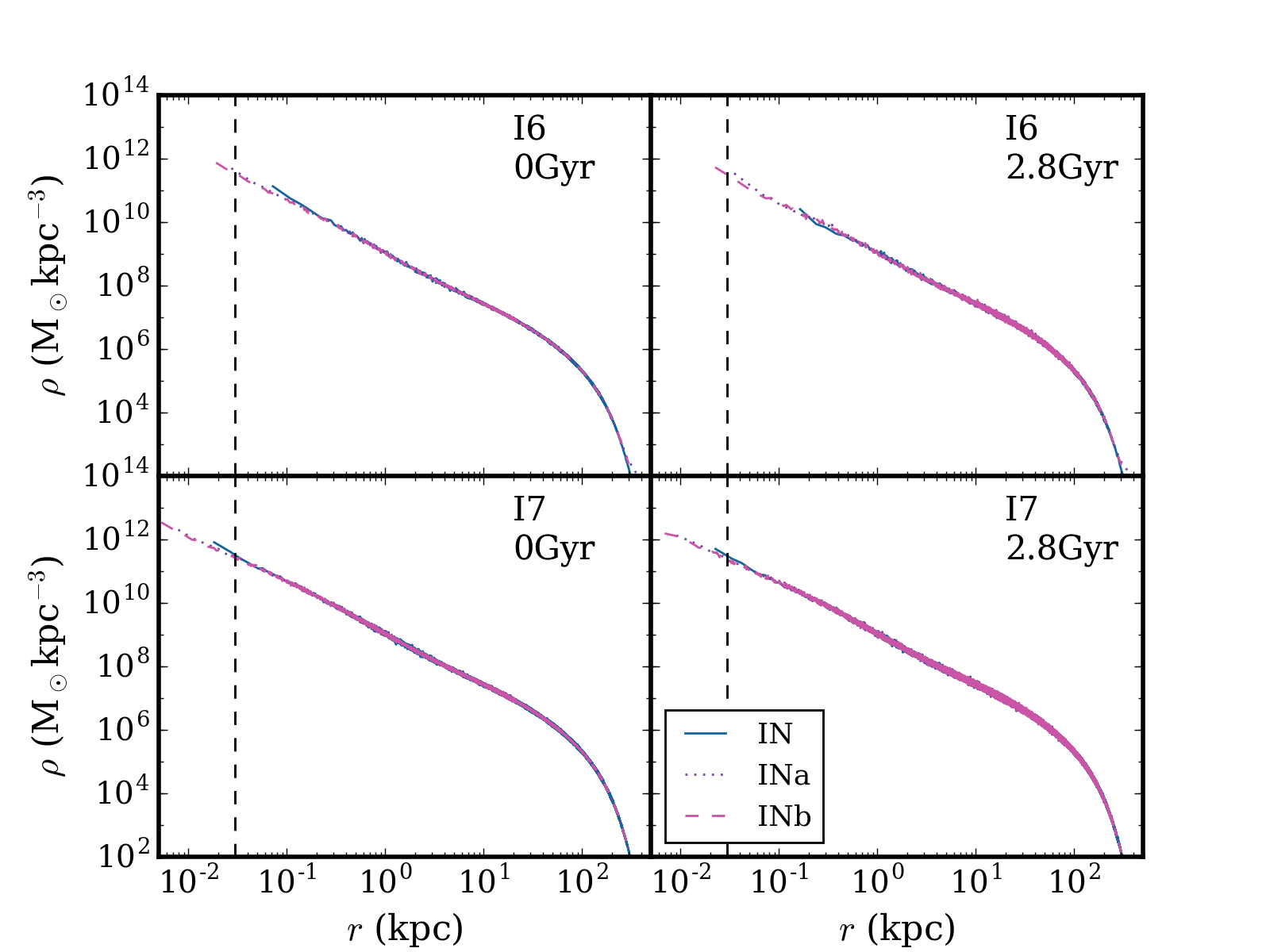}
	\caption{Spatial density profiles of the isolated models with $N=10^{6}$ particles (top panels) and $N=10^{7}$ particles (bottom panels) at time $t=0.0\gyr$ (left) and at the later time $t=2.8\gyr$ (right). The dashed vertical line represents the star-star softening length. The profiles show the same trends as the bulge particle distributions, with all the multi-mass models showing stability at the innermost radii.}
    \label{fig:INdens}
\end{figure}

\begin{figure}
\centering
	\includegraphics[width=\columnwidth]{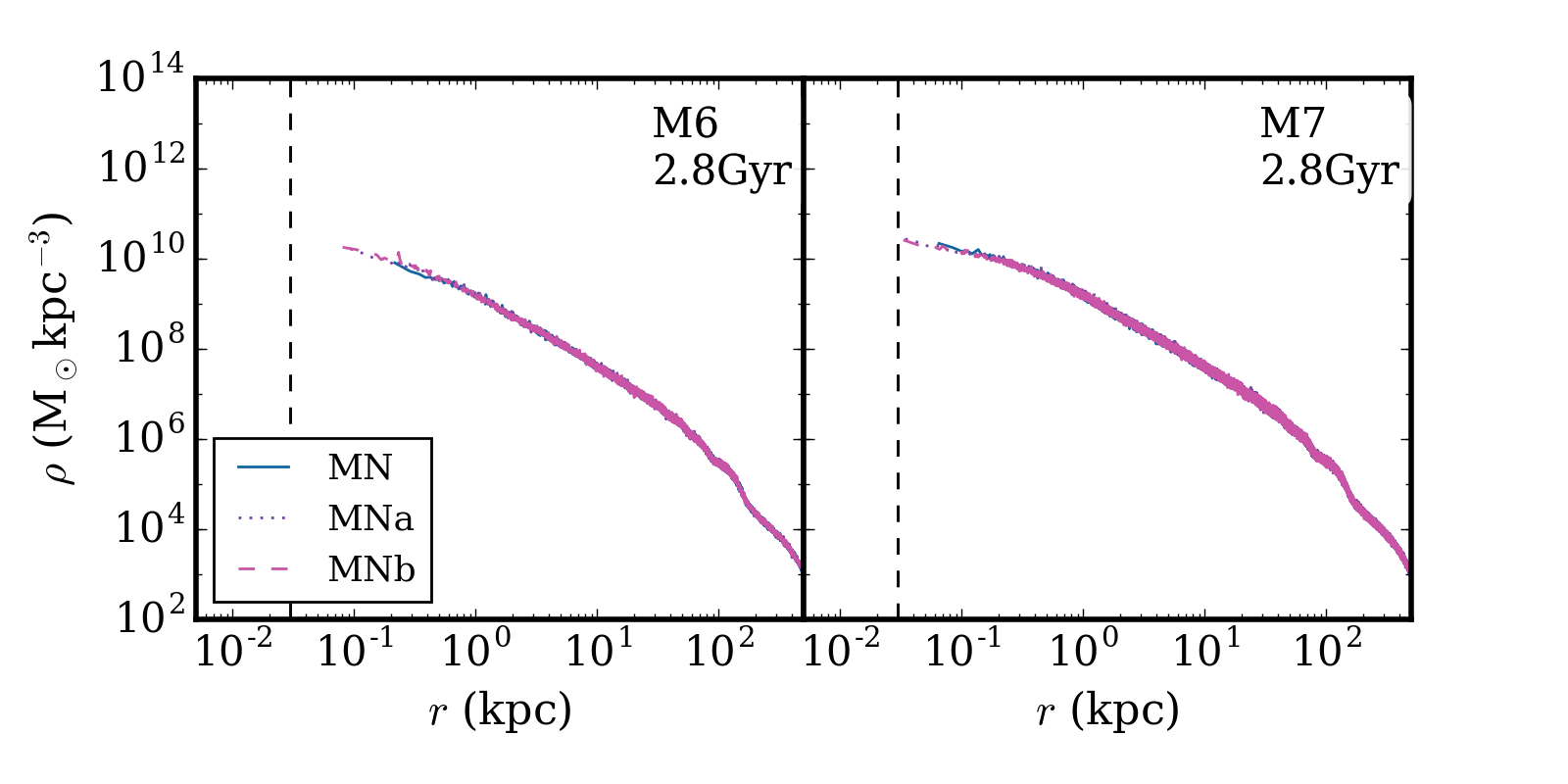}
	\caption{Spatial density profiles of the merger models with $N=10^{6}$ particles (left) and $N=10^7$ (right) at the later time $t=2.8\gyr$. The dashed vertical line represents the star-star softening length.} 
    \label{fig:MNdens}
\end{figure}

The radial density profiles shown in Fig.~\ref{fig:INdens} and \ref{fig:MNdens} support the conclusions from the particle distributions. The schemes are effective at populating the central regions of the models, and the distributions are stable over time. The flattening observed in the inner density profiles of the merger models can be attributed to core scouring during the hardening phase of the black hole binary. 


\begin{figure}
\centering
	\includegraphics[width=0.98\columnwidth]{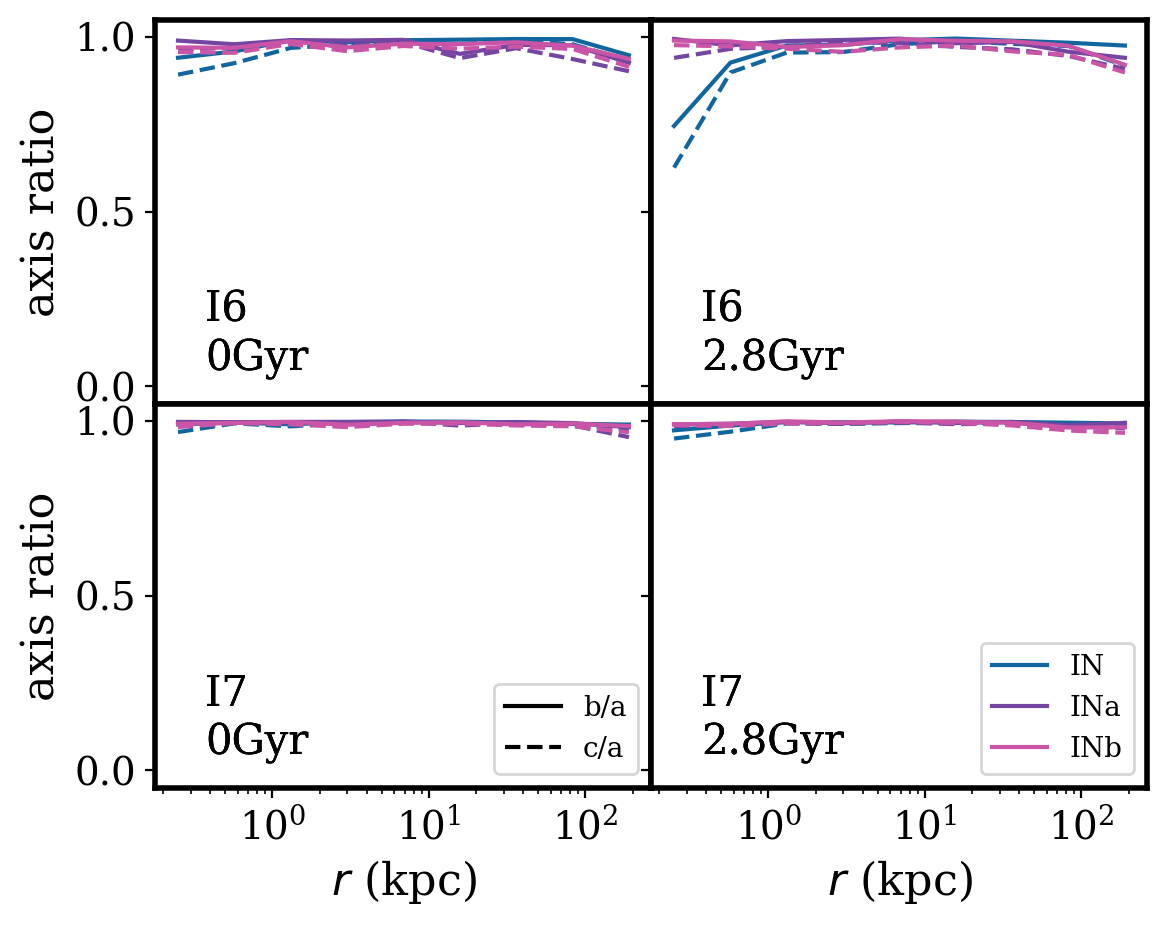}
	\caption{Axis ratios of the isolated models as a function of radius for $N=10^{6}$ particles (top) and $N=10^{7}$ particles (bottom) at time $t=0.0\gyr$ (left) and at the later time $t=2.8\gyr$ (right): $b/a$ (solid lines) and $c/a$ (dashed lines). The spherical models maintain their shape over time even when the mass refinement schemes are applied.}
    \label{fig:INbca}
\end{figure}

\begin{figure}
\centering
	\includegraphics[width=0.98\columnwidth]{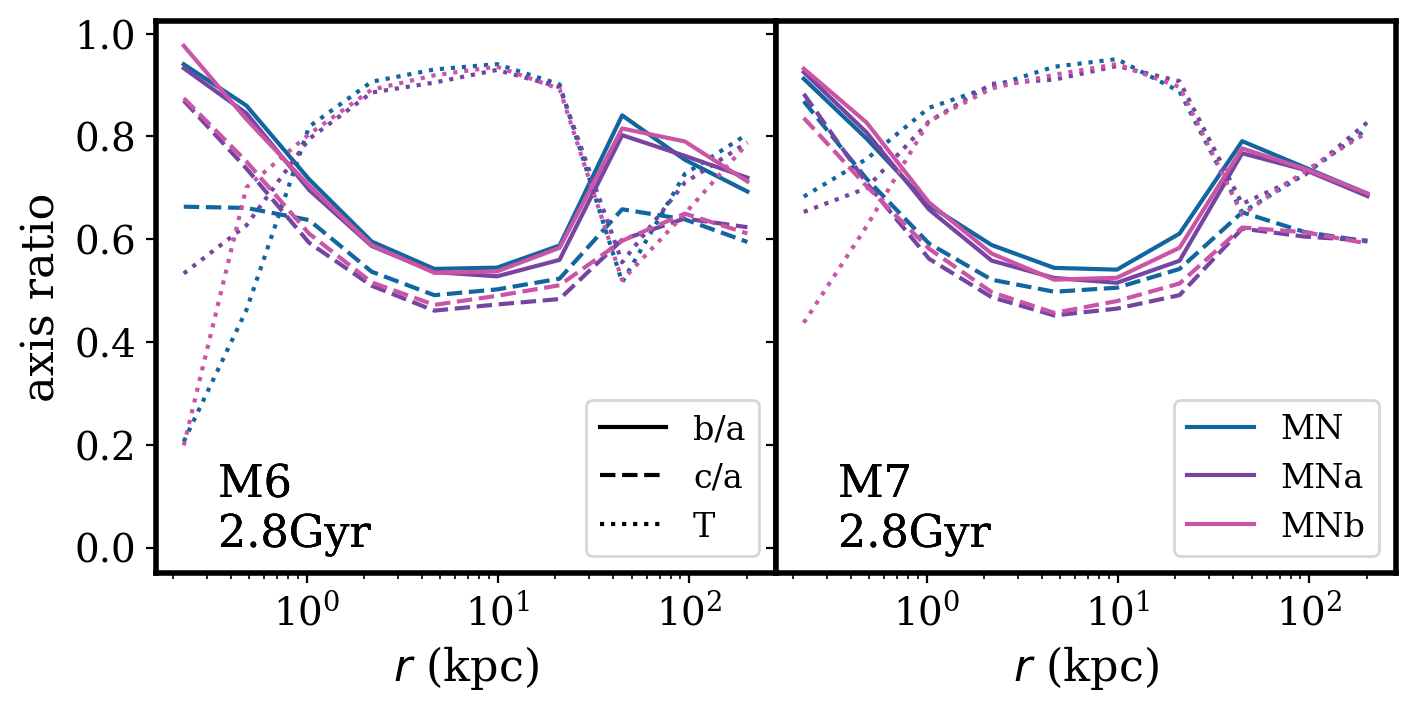}
	\caption{Axis ratios ($b/a$, solid lines) and $c/a$, dashed lines) and triaxiality parameter (dotted lines) of the merger models as function of radius with $N=10^{6}$ particles (left) and $N=10^{7}$ particles (right) at the later time $t=2.8\gyr$.}
    \label{fig:MNbca}
\end{figure}

The shape and kinematic properties of the models are also preserved by the refinement schemes. The shape of a galaxy or merger remnant can be characterised by the axis ratios of an ellipsoid fit to the stellar distribution, defined as
\begin{equation}
    1 = \left(\frac{x}{a}\right)^{2} + \left(\frac{y}{b}\right)^{2} + \left(\frac{z}{c}\right)^{2}
	\label{eq:abc}
\end{equation}
where $a>b>c>0$ are the axes of symmetry of the ellipsoid. The ratios $b/a$ and $c/a$  are equal to unity for spherical models, and decrease for axisymmetric ($b/c <1$) or triaxial ($b/c <1$ and $b/a<1$) models \citep{1985deZeeuw}. The triaxiality parameter provides a measure of the departure from sphericity, computed as
\begin{equation}
    T = \frac{1-b^{2}/a^{2}}{1-c^{2}/a^{2}}
	\label{eq:abcT}
\end{equation}
with values $T>0.5$ for prolate spheroids, $T<0.5$ for oblate spheroids, and $T=0.5$ for maximum triaxiality. We calculate the axis ratios using \textsc{Pynbody}, a Python package for analysis of astrophysical simulations \citep{2013Pontzen}.  The radial dependence of the axis ratios for the stellar component is shown in Figure~\ref{fig:INbca} for the isolated models and in Figure~\ref{fig:MNbca} for the merger models, including also the triaxiality parameter $T$.
In all isolated models, the sphericity is maintained over the simulation time, save for a noticeable departure at the innermost radii in model I6. This is likely due to low resolution in the innermost radial bins, which is resolved by the mass refinement in models I6a/b. The shape is preserved by the schemes also for the merger models, which display significant triaxiality, here in the form of  a prolate shape ($T>0.5$), in agreement with results from  \citet{Gualandris2017, 2018Bortolas}.


\begin{figure}
\centering
	\includegraphics[width=\columnwidth]{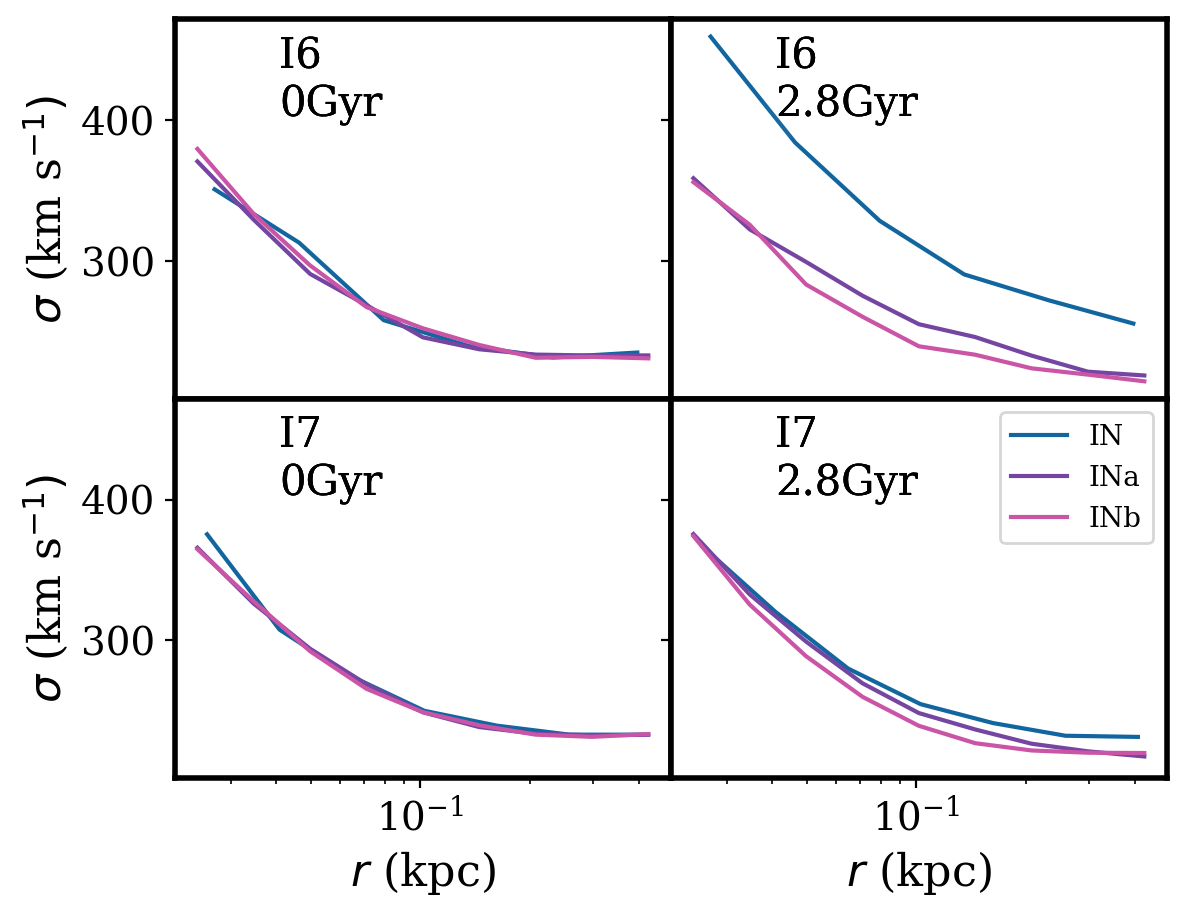}
	\caption{Stellar velocity dispersion profiles of the isolated models with $N=10^{6}$ particles (top panels) and $N=10^{7}$ particles (bottom panels) at time $t=0.0\gyr$ (left) and at the later time $t=2.8\gyr$ (right).}
    \label{fig:INvdisp}
\end{figure}

\begin{figure}
\centering
	\includegraphics[width=\columnwidth]{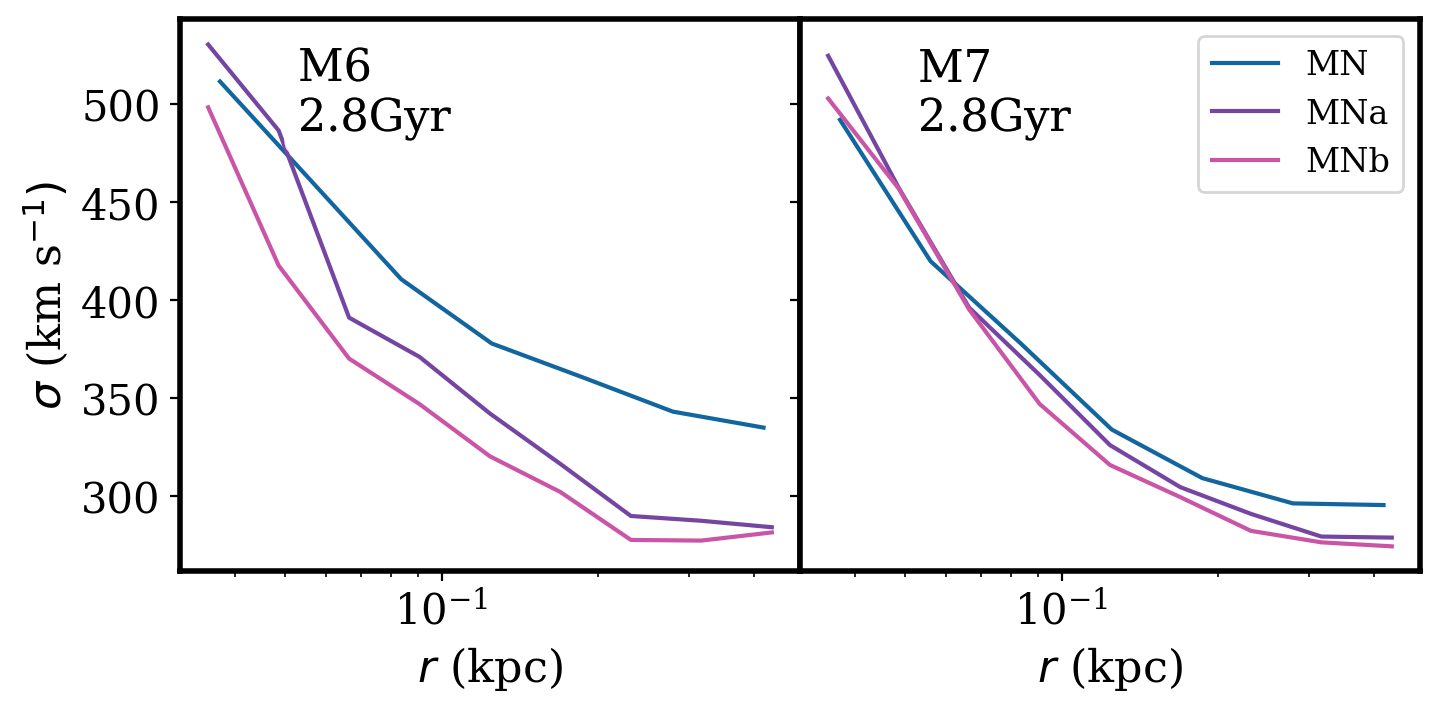}
	\caption{Stellar velocity dispersion profiles of the merger models with $N=10^{6}$ particles (left) and $N=10^{7}$ particles (right) at the later time $t=2.8\gyr$.}
    \label{fig:MNvdisp}
\end{figure}

We present 3D stellar velocity dispersion profiles in Fig.~\ref{fig:INvdisp} for the isolated models and Fig.~\ref{fig:MNvdisp} for the merger models, showing a small degree of deviation between the low and high resolution models. In particular, models I6 and M6 show a higher velocity dispersion at all radii. The mass refined models, on the other hand, are consistent at both resolutions, and consistent with M7.

\begin{figure}
\centering
	\includegraphics[width=1.\columnwidth]{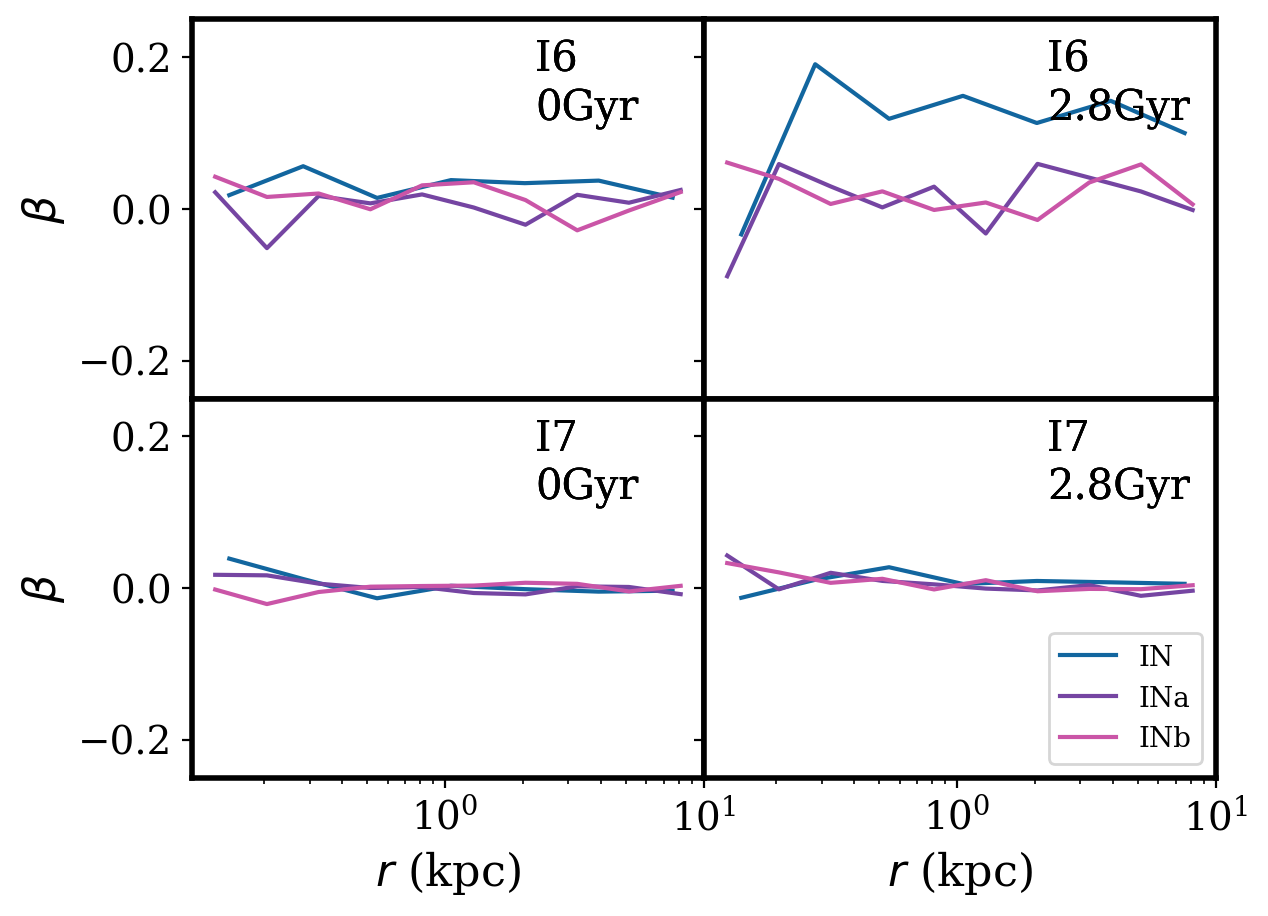}
	\caption{Stellar velocity anisotropy profile of the isolated models with $N=10^{6}$ particles (top panels) and $N=10^{7}$ particles (bottom panels) at time $t=0.0\gyr$ (left) and at later time $t=2.8\gyr$ (right). All models maintain an isotropic profile over time, except for I6 which shows a moderate tangential anisotropy at late times, likely due to low resolution.}
 
    \label{fig:betaIN}
\end{figure}

\begin{figure}
\centering
	\includegraphics[width=1.\columnwidth]{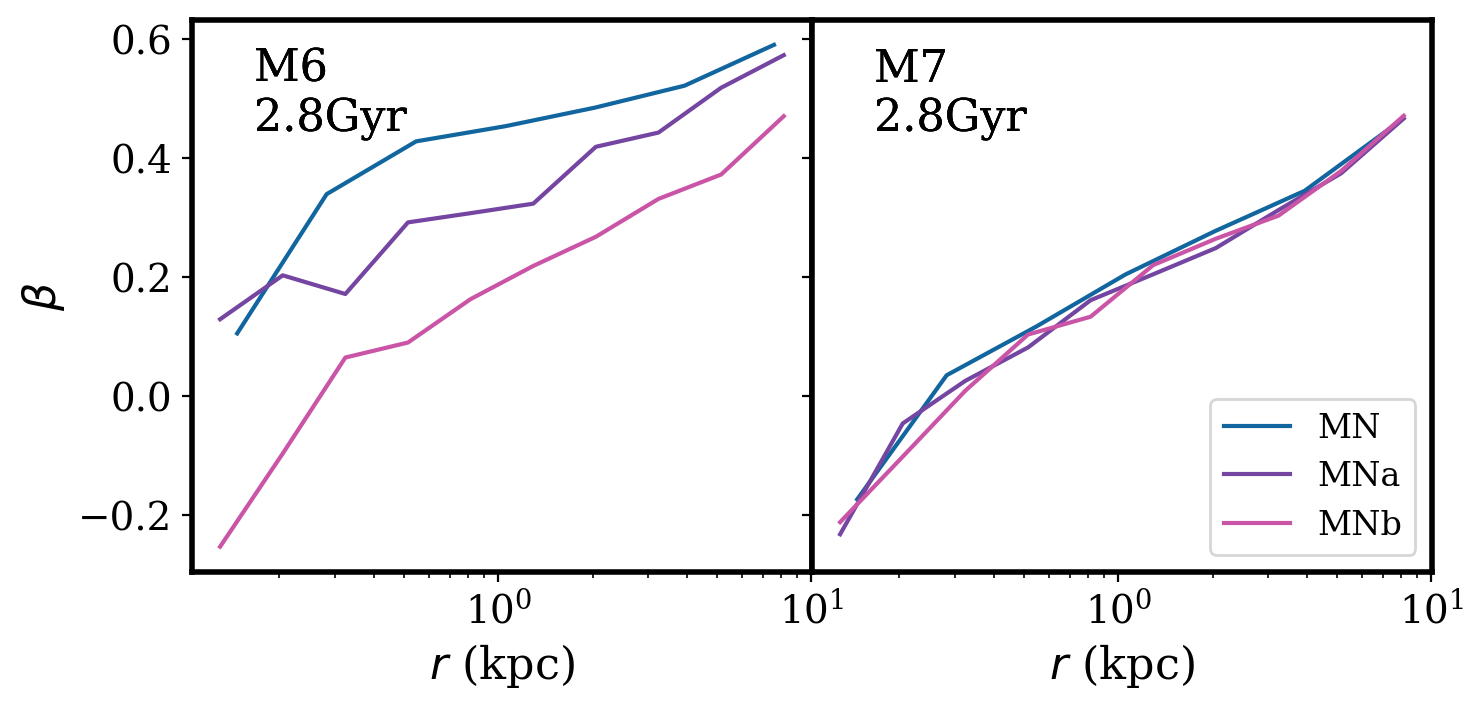}
	\caption{Stellar velocity anisotropy profile of the merger models with $N=10^{6}$ particles (left) and $N=10^{7}$ particles (right) at the later time $t=2.8\gyr$. A modest discrepancy between the models is observed at the lower resolution, while at the higher resolution the models are indistinguishable.}
    \label{fig:betaMN}
\end{figure}

A similar behaviour can be seen in the anisotropy profiles (see Fig.~\ref{fig:betaIN} and ~\ref{fig:betaMN}), calculated according to \citep{2008BT}
\begin{equation}
    \beta(r) = 1 - \frac{\sigma_{\theta}^{2}(r)}{\sigma_{r}^{2}(r)}
	\label{eq:beta}
\end{equation}
where $\sigma_{\theta}(r)$ is the tangential velocity dispersion and $\sigma_{r}(r)$ is the radial velocity dispersion. The isolated models are isotropic at all radii, with the exception of model I6 which displays a small radial bias over time. The merger models display the expected anisotropy with an inner tangential bias, imparted from core scouring as stellar scatterings during the hardening phase of binary evolution preferentially eject stars on radial orbits \citep{1997Quinlan,2014Thomas}.

Interestingly, there is a modest discrepancy at low resolution even when the schemes are applied, suggesting that the binary hardening phase is sensitive to the specific sequence and properties of encounters with stars experienced over time. No discrepancy is observed at the larger resolution, with or without mass refinement.
 

\begin{figure}
\centering
	\includegraphics[width=\columnwidth]{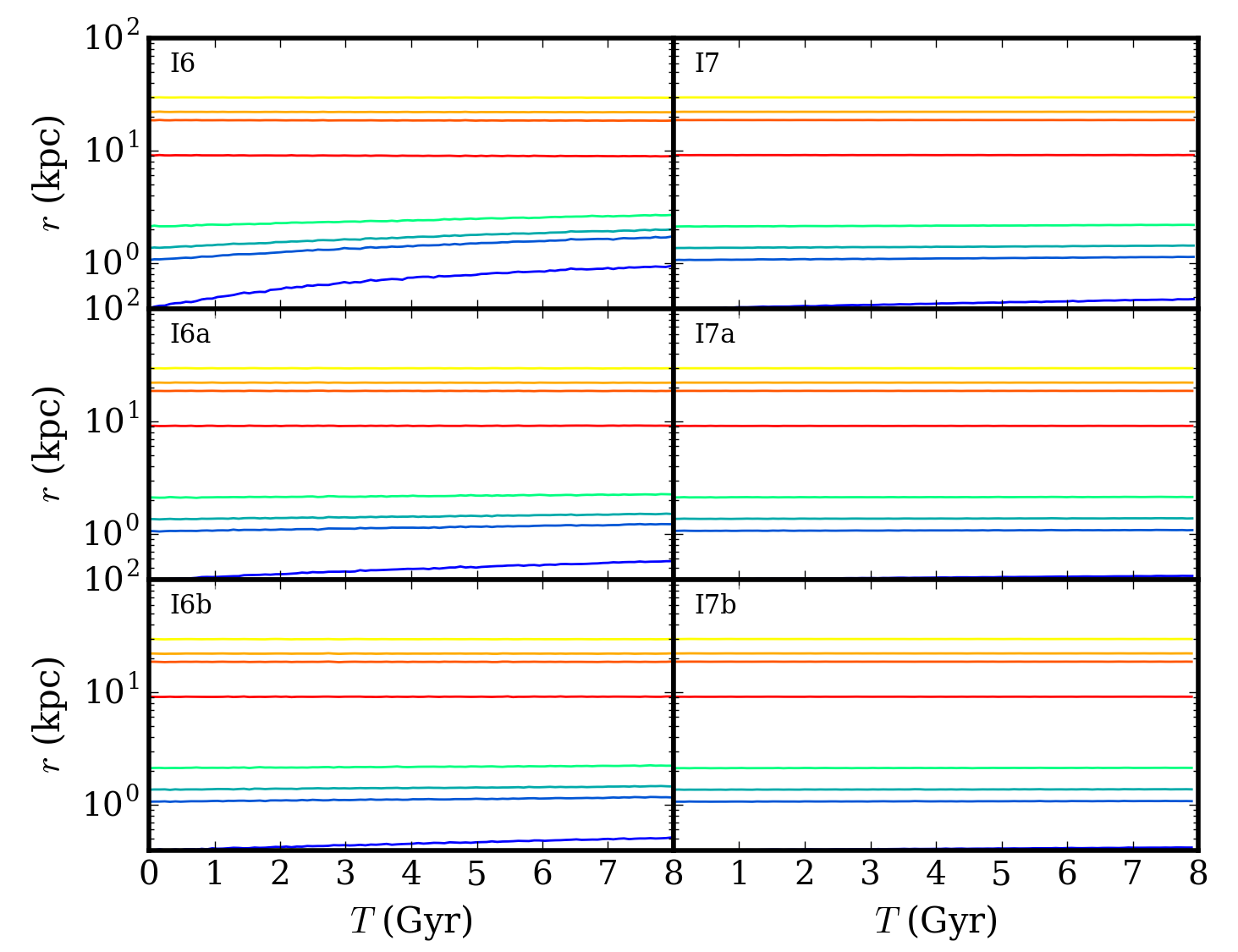}
	\caption{Lagrangian radii for bulge (bottom four curves) and halo (top four curves) particles for all isolated models with $N=10^{6}$ (left) and $N=10^{7}$ (right) across the simulation time.  The curves correspond to radii containing  mass fractions of 3\%, 10\%, 13\%, and 20\% of the total bulge mass. I6 shows a significant expansion at the lowest bulge radius, which is notably reduced in the multi-mass models. The halo particles remain stable in all models.}
    \label{fig:LagIN}
\end{figure}

\begin{figure}
\centering
	\includegraphics[width=\columnwidth]{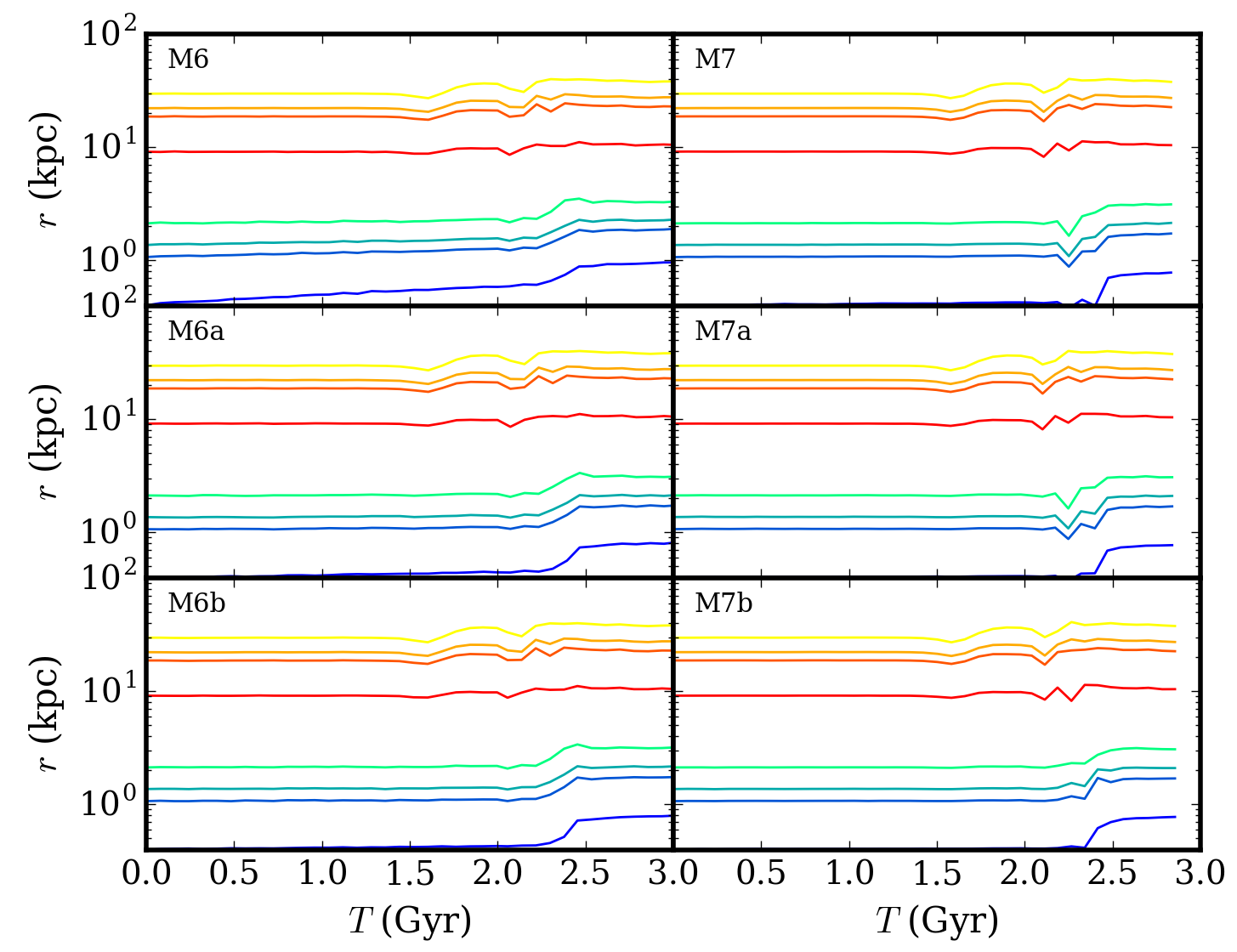}
	\caption{Lagrangian radii for bulge (bottom four curves) and halo (top four curves) particles for all merger models 
	 with $N=10^{6}$ (left) and $N=10^{7}$ (right) across the simulation time.  The curves correspond to radii containing  mass fractions of 3\%, 10\%, 13\%, and 20\% of the total bulge mass. The strong disturbance at $t\sim2.5\gyr$ marks the completion of the merger process. The M6 model shows a significant expansion at the smallest radii, similarly to I6.}
    \label{fig:LagMN}
\end{figure}

The Lagrangian radii shown in Fig.~\ref{fig:LagIN} and Fig.~\ref{fig:LagMN} for isolated and merger models, respectively, reveal a small expansion in the bulge particles, and predominantly those at lower radii. This owes to relaxation effects at small radii and at low resolution, as indicated by the dependence of the effect on particle number and radius. The expansion lessens with increasing radius, due to an increase in relaxation time with radius, and is almost negligible in both the refined $N=10^6$ models and all the $N=10^7$ models. Furthermore, it does not affect the halo particles. The strong variation at $t=2.0-2.5\gyr$ in the merger models is a signature of the merger process, and can be taken as a measure of the merger time.

\begin{figure}
\centering
	\includegraphics[width=\columnwidth]{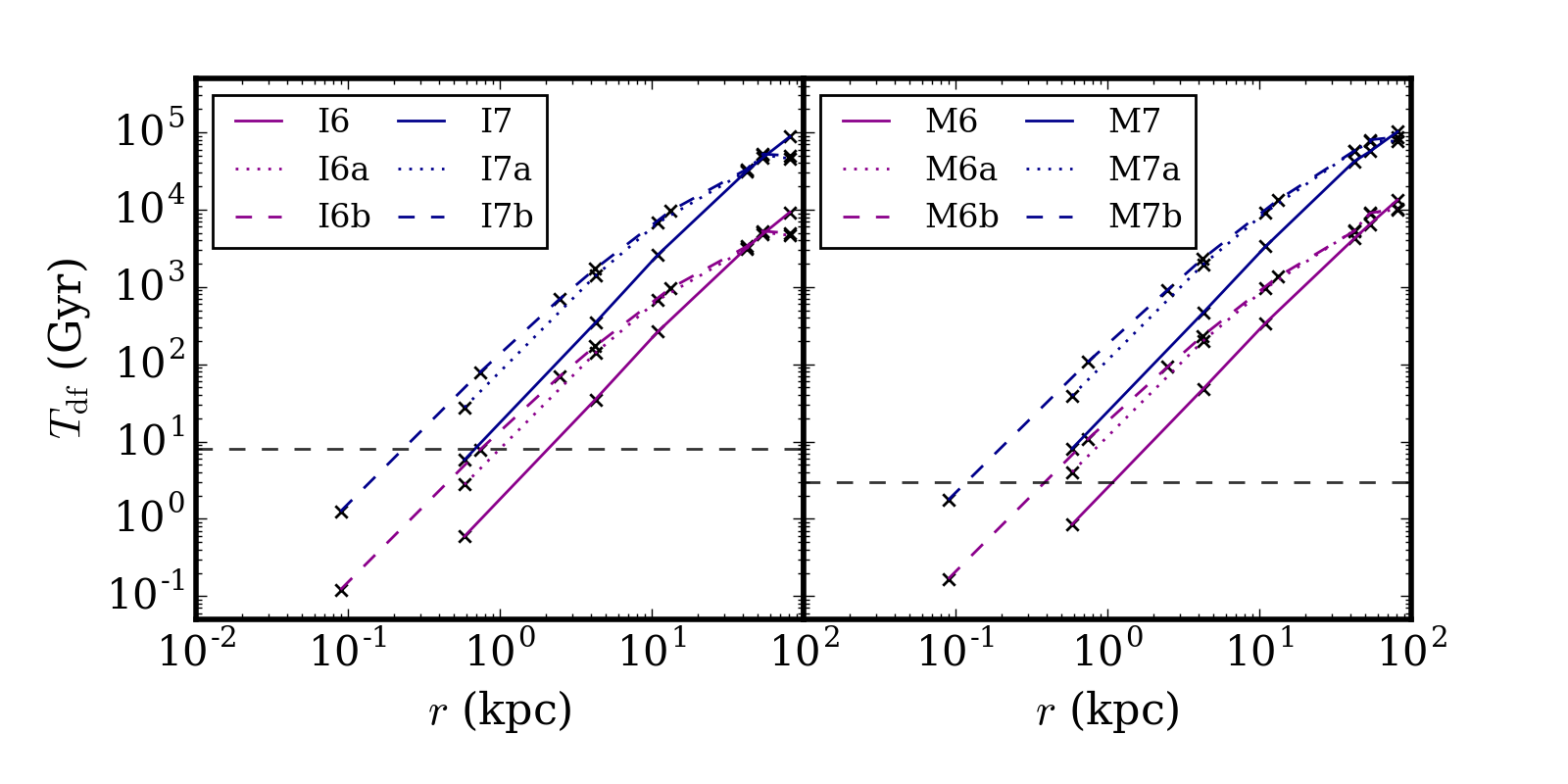}
	\caption{Dynamical friction timescale of the halo particles for all models at time $t=2.8\gyr$, for the isolated models (left) and the post-merger models (right). The dashed horizontal lines mark the duration of the simulations (black). The points mark the centres of each radial bin in the multi-mass schemes. The particles are affected by dynamical friction at $r\lesssim2\kpc$ in I6, and at $r\lesssim1\kpc$ in the multi-mass models. In model I7, dynamical friction only affects particles within the central $\lesssim 500\pc$, which is reduced to the central $\lesssim 200\pc$ in the multi-mass models. The multi-mass models show a downturn in $\tdf$ at the outer radii, where the particle masses are most increased. The merger models show similar trends to the isolated models.}
    \label{fig:tdynI}
\end{figure}

The results presented in Fig.~\ref{fig:INdens} and \ref{fig:MNdens} are in agreement with expectations based on a simple calculation of the dynamical friction timescale for point mass particles. A particle of mass $M$ in a system of lighter particles will sink from a radius $r_i$ on a timescale given by Chandrasekhar's dynamical friction formula \citep{2008BT}
\begin{equation}
    \tdf = \frac{19\gyr}{5.8}\frac{\sigma}{200\kms}\frac{r_{i}}{5\kpc}^{2}\frac{10^{8}\msun}{M}
	\label{eq:tdf}
\end{equation}
where $\sigma$ is the local stellar velocity dispersion.  The timescale $\tdf$ of the halo particles is plotted in Fig.~\ref{fig:tdynI} for both isolated and merger models as a function of radial distance. The chosen radial positions correspond to the centres of the scheme zones. The horizontal line indicates the total simulation time of $8\gyr$ for the isolated models, and $4\gyr$ for the merger models. Halo particles in the I6 models are affected by dynamical friction against the sea of bulge particles at distances $r \lesssim 2\kpc$ while mass refined models I6a and I6b are only affected within $r \lesssim 1\kpc$. At higher resolution, sinking is expected in model I7 only for particles within the central $\lesssim 500\pc$, which is reduced to the central $\lesssim 200\pc$ in the mass refined models I7a and I7b. We notice that the multi-mass models experience a downturn in the calculated $\tdf$ at large radii $r\gtrsim 10\kpc$, due to the increase in particle masses introduced by the schemes to compensate for the reduced particle number. Even so, the schemes appear beneficial out to $\sim40\kpc$. The merger models exhibit a very similar behaviour to the isolated models.

We note that the total wall-clock time of the refined and unrefined models is comparable, and the mass refinement schemes do not produce any appreciable slow-down in our Griffin runs.

\section{Evolution of the massive black hole binaries}
\label{sec:orbev}
We follow the evolution of the MBHs through the merger, the formation of a bound binary and the collisionless losscone refilling phase and employ a semi-analytical model to compute the merger timescale for the different models due to GW emission.  We examine the effect of resolution and of the multi-mass schemes on the orbital elements of the binary as well as the coalescence timescale. Using the 7 random realisations of the $N=10^{6}$ merger  models, we quantify the scatter in the orbital elements of the binary and the total merger time. We adopt the dispersion in semi-major axis, eccentricity and merger timescales as our criterion for convergence.

\begin{figure*}
\centering
	\includegraphics[width=2\columnwidth]{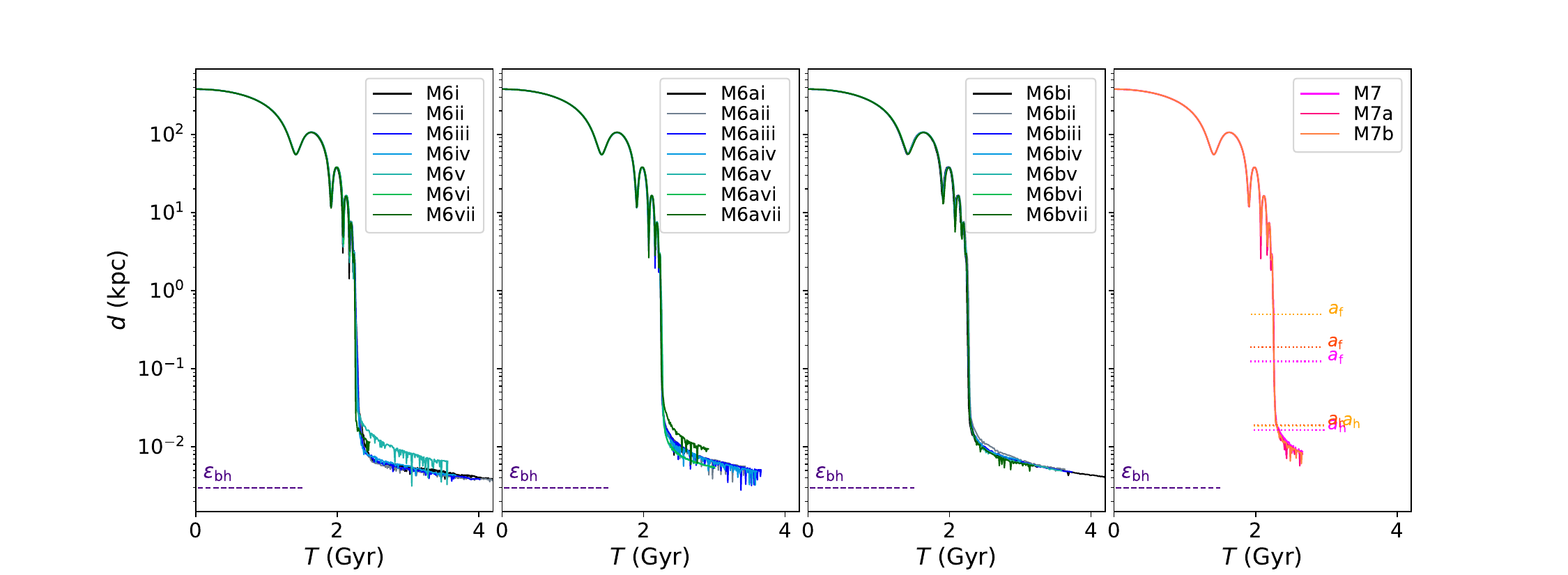}
	\caption{The evolution of the distance $d$ between the MBHs for all models at the lower (left) and higher (right) resolution. Overall, all models consistently reproduce the evolution of $d$ over time. However the M6 and M6a models show small discrepancies at late times, which disappear in M6b and at $N=10^7$. The dashed horizontal line marks the softening length for the BH-BH and BH-particle interactions. The dotted lines in the rightmost panel mark the critical separations of the binary: $a_{f}$, approximately where the dynamical friction phase ends, and $a_{h}$, the hard binary separation.}
    \label{fig:bh_dist}
\end{figure*}

\begin{figure*}
\centering
	\includegraphics[width=2\columnwidth]{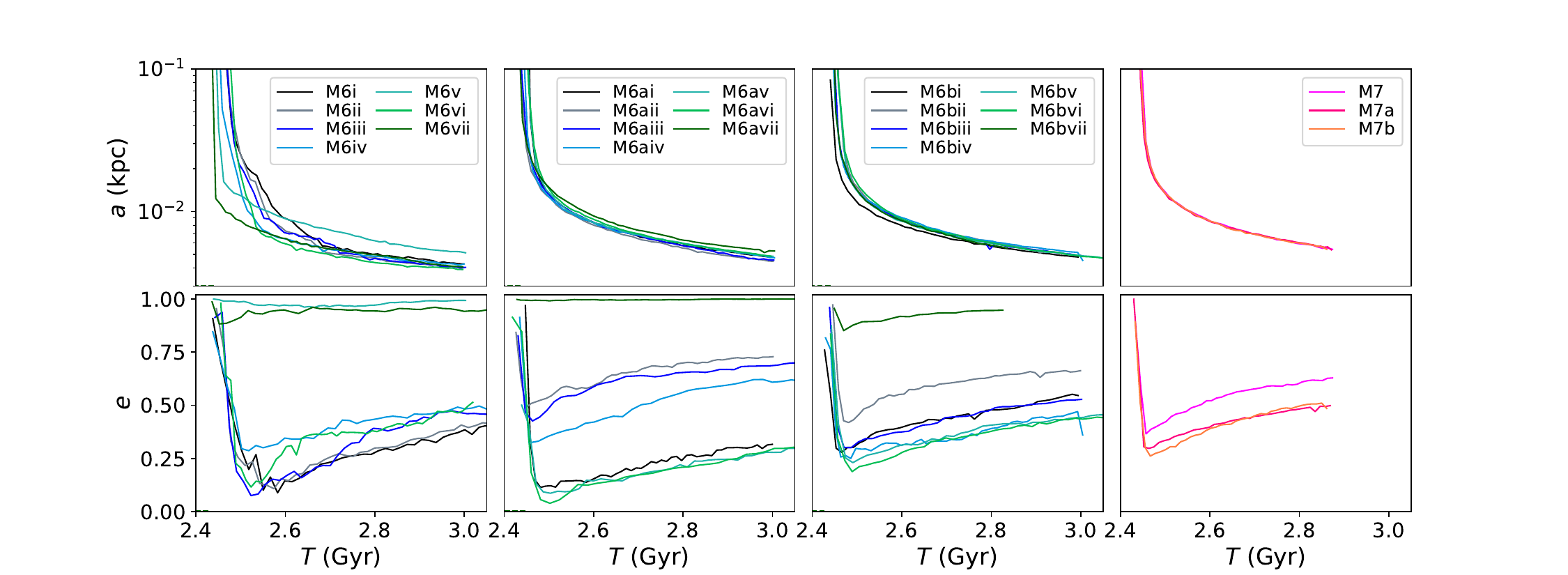}
	\caption{The evolution of the semi-major axis $a$ (top) and the eccentricity $e$ (bottom) of the BHB with time, computed from the \textsc{griffin} simulations, for all models. The lower resolution models show a larger scatter in semi-major axis and especially in eccentricity with respect to the higher resolution models. At $N=10^6$, the refinement schemes prove to be effective at reducing the spread in the orbital elements. Model M6b is the most consistent with the higher resolution models. In addition, we find a clear trend of reduced scatter in the semi-major axis from left to right, with models M7/M7a/M7b showing excellent agreement. The same trend is seen in the eccentricity, though clearly with a larger scatter, due to stronger stochasticity.}
     \label{fig:bh_orb}
\end{figure*}

\begin{figure*}
\centering
\includegraphics[width=2\columnwidth]{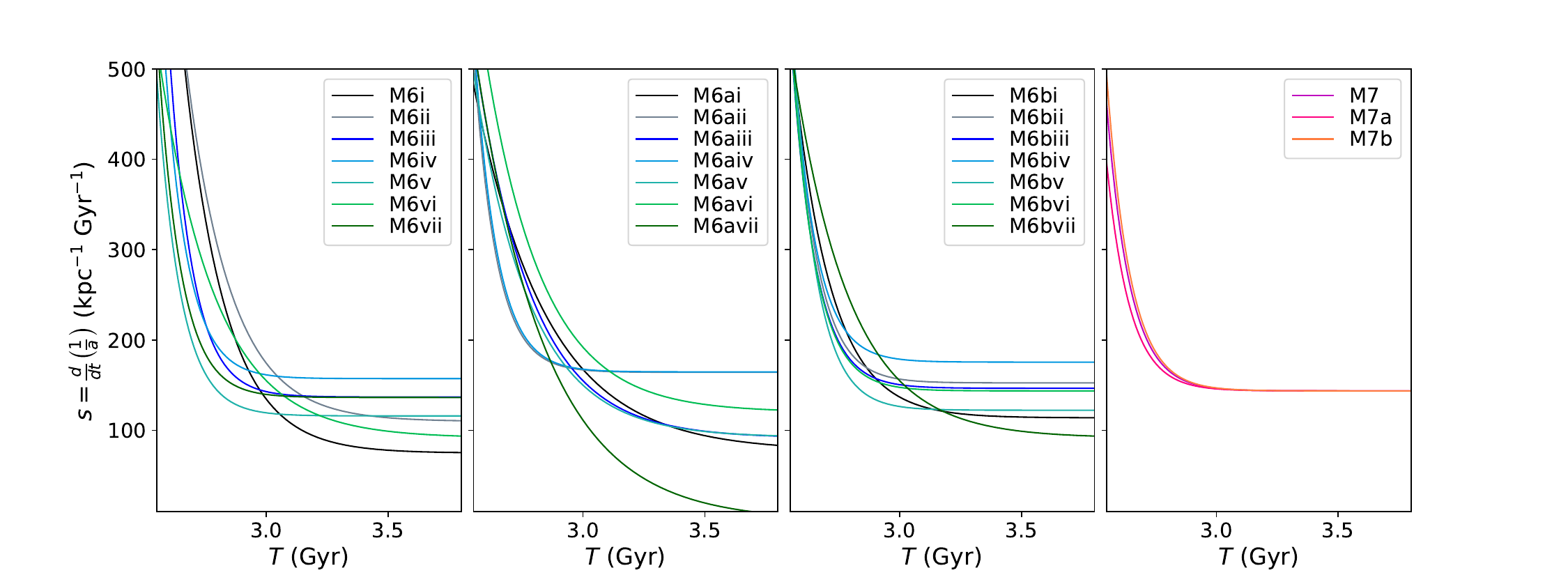}
	\caption{Evolution of the hardening rate, computed numerically as in Eq.~\ref{eq:s_t}, for all models. The scatter decreases with increasing resolution, from left to right.}
    \label{fig:hard_rate}
\end{figure*}

The three characteristic phases of binary evolution can clearly be seen in Fig.~\ref{fig:bh_dist}, which shows the distance between the MBHs as a function of time. The galaxies first evolve under the effects of dynamical friction, which brings them together during the merger and, at the same time, causes the MBHs to sink to the centre of the merger remnant, where 
they form a pair and eventually a bound binary. At the end of this phase, the MBHs are approximately at a separation $\af$, where the enclosed stellar mass is equal to twice the mass of the secondary MBH, as
\begin{equation}
M_{*}(\af) = 2\,M_{2}.
\label{eq:af}
\end{equation}
Three-body interactions between the binary and background stars become increasingly important until they dominate the evolution, as the influence of dynamical friction wanes once the binary begins to reach thermal equilibrium with surrounding stars \citep{Antonini2012,Kelley2017}. Such interactions are known as slingshot interactions as the stars remove energy and angular momentum from the binary and are ejected to large distances. As a result, the MBHs spiral closer together \citep{Begelman1980,2013MerrittB} until all stars initially belonging to the losscone are ejected; this occurs roughly at the {\it hard-binary} separation $\ah$, which is of order a parsec for black holes of $M\sim10^8\msun$. There are different definitions of the hard-binary separation in the literature, here we adopt the formal one given by the separation 
where the relative velocity of the binary surpasses the velocity dispersion of local stars \citep{2013MerrittB} 
\begin{equation}
    a_{h} = \frac{G\mu}{4\sigma^{2}},
	\label{eq:ah}
\end{equation}
where $\mu=M_{1}M_{2}/(M_{1}+M_{2})$ is the reduced mass of the binary. Beyond $\ah$, binary hardening relies on collisionless repopulation of the losscone, until GW emission becomes important (at roughly milliparsec scales) and drives the MBHs to coalescence.
Our numerical simulations are terminated when the separation between the MBHs reaches $\epsilon_{\rm BH}$, the BH softening length, which we have set to be smaller than $\ah$ to ensure that the evolution is followed self-consistently well into the losscone refilling phase.

Fig.~\ref{fig:bh_dist} shows that all models reproduce the same evolution for the separation between the MBHs all the way down to the BH softening length. The same is not true for the evolution of the orbital elements of the binary (see Fig.~\ref{fig:bh_orb}), which are characterised by a significant spread among the different random realisations at the lower resolution.  The spread decreases with increasing resolution (from left to right), though much more slowly for the eccentricity. This appears to be more sensitive to perturbations and stochastic effects, as may be expected based on the results of \citet{Nasim2020, Gualandris2022, Rawlings2023}. In particular, a few random realisations show binaries forming with very large eccentricity, despite a moderate initial orbital eccentricity for the galactic merger. At the highest resolution, model M7 shows a non negligible difference in eccentricity with respect to models M7a/b, implying that mass refinement is required even at $N=10^7$ to reduce the scatter in eccentricity. Because the coalescence timescale of BHBs due to GW emission is strongly dependent on the orbital elements, and in particular on the eccentricity, we expect that these results will affect the time to coalescence.

We follow the late evolution of the binaries with a semi-analytical model that solves the coupled differential equations for the orbital elements under the effects of both stellar hardening and GW emission \citep{Gualandris2022}
\begin{equation}
    \frac{da}{dt} = \left.\frac{da}{dt}\right|_{*} + \left.\frac{da}{dt}\right|_{GW}
	\label{eq:samorba}
\end{equation}
\begin{equation}
    \frac{de}{dt} = \left.\frac{de}{dt}\right|_{*} + \left.\frac{de}{dt}\right|_{GW}
	\label{eq:samorbe}
\end{equation}
where the first term represents the contribution from three-body stellar interactions and the second term models GW emission. 

The contribution from the stellar interactions can be written as
\begin{equation}
    \left.\frac{da}{dt}\right|_{*} = -s(t)a^{2}
	\label{eq:3Ba}
\end{equation}
\begin{equation}
    \left.\frac{de}{dt}\right|_{*} = s(t)Ka
	\label{eq:3Be}
\end{equation}
where $s$ represents the time-dependent hardening rate of the binary
\begin{equation}
    s(t) = \frac{d}{dt}\left(\frac{1}{a}\right)
	\label{eq:s_t}
\end{equation}
while $K$ is the eccentricity growth rate
\begin{equation}
    K = \frac{de}{d~\textrm{ln}(1/a)}
	\label{eq:k_t}
\end{equation}
as defined by \citet{1996Quinlan}.

The contribution from GW emission can be derived for two point masses directly from the Einstein field equations in the non relativistic limit \citep{peters1964}
\begin{equation}
    \left.\frac{da}{dt}\right|_{GW} = -\frac{64}{5}\frac{G^{3}M_{1}M_{2}M_{T}}{c^{5}a^{3}(1-e^{2})^{7/2}}\left( 1+\frac{73}{24}e^{2}+\frac{37}{96}e^{4} \right)
	\label{eq:petersa}
\end{equation}
\begin{equation}
    \left.\frac{de}{dt}\right|_{GW} = -\frac{304}{15}e\frac{G^{3}M_{1}M_{2}M_{T}}{c^{5}a^{4}(1-e^{2})^{5/2}}\left( 1+\frac{121}{304}e^{2} \right)
	\label{eq:peterse}
\end{equation}
where $M_{T}=M_{1}+M_{2}$ is the total mass of the binary.

\begin{figure*}
\centering
 \includegraphics[width=2\columnwidth]{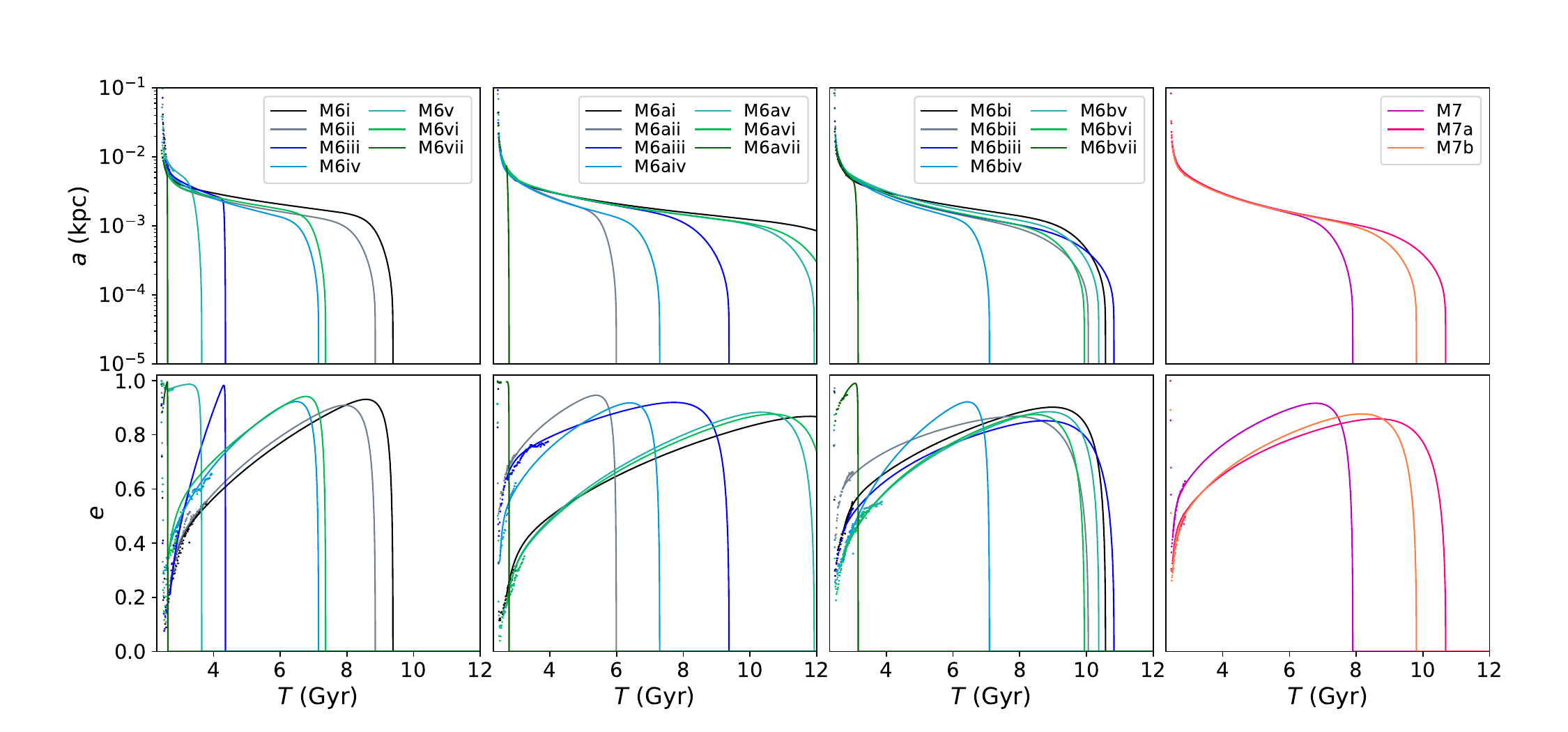}
	\caption{The evolution of the semi-major axis $a$ (top) and the eccentricity $e$ (bottom) of the binary with time, for all models. The evolution simulated with \textsc{griffin} is shown with points, whilst the subsequent evolution modelled with the semi-analytic model including GWs is shown with solid lines. 
 }
    \label{fig:sam}
\end{figure*}

\begin{table*}
    \centering
    \caption{Characteristic timescales for the binaries throughout the evolution.  From left to right: the simulation identifier, the time to reach $a_{f}$ (an estimate of the time spent in the dynamical friction phase), the time spent in stellar hardening, the time spent in GW emission, the total time from the start of the simulation to BHB merger, the dispersion of the eccentricity at the binary formation time, the dispersion of the semi-major axis and the dispersion of the binary timescale.}
    \label{tab:BHmergers}
    \begin{tabular}{l|ccccccc}
    \hline
    Scheme & $\taf\,(\gyr)$ & $T_{\rm hard}\,(\gyr)$ & $T_{\rm GW}\,(\gyr)$ & $T_{\rm merg}\,(\gyr)$ & $\sigma_{e}$ & $\sigma_{a}$ & $\sigma_{m}$\\ 
    \hline
    M6i    &  $2.27$   & $5.73$ & $1.38$ & $9.38$ & $0.34$ & $0.00112$ & $2.51$ \\
    M6ii     &  $2.27$   & $5.14$ & $1.44$ & $8.85$ & $0.34$ & $0.00112$ & $2.51$ \\
    M6iii     &  $2.27$   & $1.96$ & $1.34$ & $4.35$ & $0.34$ & $0.00112$ & $2.51$ \\
    M6iv     &  $2.27$   & $3.87$ & $1.01$ & $7.15$ & $0.34$ & $0.00112$ & $2.51$ \\
    M6v     &  $2.27$   & $0.63$ & $0.37$ & $3.28$ & $0.34$ & $0.00112$ & $2.51$ \\
    M6vi     &  $2.27$   & $4.07$ & $1.02$ & $7.36$ & $0.34$ & $0.00112$ & $2.51$ \\
    M6vii     &  $2.27$  & $0.33$ & $0.02$ & $2.62$ & $0.34$ & $0.00112$ & $2.51$ \\
    M6ai    &  $2.27$   & $8.71$ & $2.98$ & $13.9$ & $0.29$ & $0.00037$ & $3.82$ \\
    M6aii    &  $2.27$   & $2.86$ & $8.62$ & $5.99$ & $0.29$ & $0.00037$& $3.82$ \\
    M6aiii    &  $2.27$   & $5.26$ & $2.33$ & $9.89$ & $0.29$ & $0.00037$& $3.82$ \\
    M6aiv    &  $2.27$   & $3.89$ & $1.25$ & $7.41$ & $0.29$ & $0.00037$& $3.82$ \\
    M6av    &  $2.27$   & $7.41$ & $2.24$ & $11.92$ & $0.29$ & $0.00037$& $3.82$ \\
    M6avi    &  $2.27$   & $8.15$ & $2.92$ & $13.31$ & $0.29$ & $0.00037$& $3.82$ \\
    M6avii    &  $2.27$   & $0.51$ & $0.0$ & $2.77$ & $0.29$ & $0.00037$ & $3.82$ \\
    M6bi    &  $2.27$   & $5.13$ & $1.06$ & $9.00$ & $0.21$ & $0.00028$ & $1.16$ \\
    M6bii    &  $2.27$   & $3.99$ & $1.48$ & $7.75$ & $0.21$ & $0.00028$ & $1.16$ \\
    M6biii    &  $2.27$   & $6.21$ & $2.35$ & $10.8$ & $0.21$ & $0.00028$ & $1.16$ \\
    M6biv    &  $2.27$   & $3.85$ & $0.98$ & $7.09$ & $0.21$ & $0.00028$ & $1.16$ \\
    M6bv    &  $2.27$   & $5.83$ & $1.88$ & $9.98$ & $0.21$ & $0.00028$ & $1.16$ \\
    M6bvi    &  $2.27$   & $5.75$ & $1.91$ & $9.93$ & $0.21$ & $0.00028$ & $1.16$ \\
    M6bvii    &  $2.27$   & $7.19$ & $1.68$ & $3.15$ & $0.21$ & $0.00028$ & $1.16$ \\
    \hline
    M7     &  $2.27$   & $4.19$ & $1.44$ & $7.89$ & - & - \\
    M7a    &  $2.27$   & $6.08$ & $2.33$ & $10.1$ & - & - \\
    M7b    &  $2.27$   & $5.53$ & $2.01$ & $9.81$ & - & - \\
    \hline
    \end{tabular}
\end{table*}

We compute the time-dependent numerical hardening rate in the hardening phase simulated with \textsc{griffin} and fit an exponential decay over time. The resulting fits are shown in Fig.~\ref{fig:hard_rate}. We observe some variation in hardening rate in the M6 models, which is significantly reduced in the M6b models.
This can explain the differences in the evolution of the semi-major axis in the M6 models. The hardening rate is fully converged at the larger resolution, with no discernible difference past 3 Gyr.

The eccentricity growth rate is approximately constant over the simulated hardening phase in all cases. Therefore, for each model, we adopt its average over the hardening phase to be used in Eqs.~\ref{eq:3Ba} and \ref{eq:3Be}, together with the fitted hardening rate.

We start the semi-analytical integration from time $2.6\gyr$ for the M6 models and from time $2.5\gyr$ for the M7 models, to ensure consistency with the $N$-body integrations, and calculate the orbital elements until coalescence, as determined by the numerical solver. In order to reduce noise, we take an average of the semi-major axis and eccentricity values within $\pm50\myr$ of the starting time.

The resulting evolution of the binary orbital elements is shown in Fig.~\ref{fig:sam} for all models.  
The predicted trajectories, shown by the solid lines, are consistent with the prior evolution from the \textsc{griffin} integrations, shown by the points. We note that stochasticity in the $N$-body evolution significantly affects the extrapolations to late times, as both three-body encounters and GW emission depend strongly on the semi-major axis and, especially, the eccentricity. Given that the hardening rate and eccentricity growth rate are very similar in all models except the M6s, the merger timescales are primarily set by the value of eccentricity at binary formation. For example, among the M6a models, the binary in model M6avi is the least eccentric and the one with the longest merger timescale. Overall, however, the refinement schemes are effective at reducing the scatter in eccentricity, with models M6b showing the smallest scatterand merger timescales more consistent with the M7 models.

\begin{figure}
\centering
	\includegraphics[width=\columnwidth]{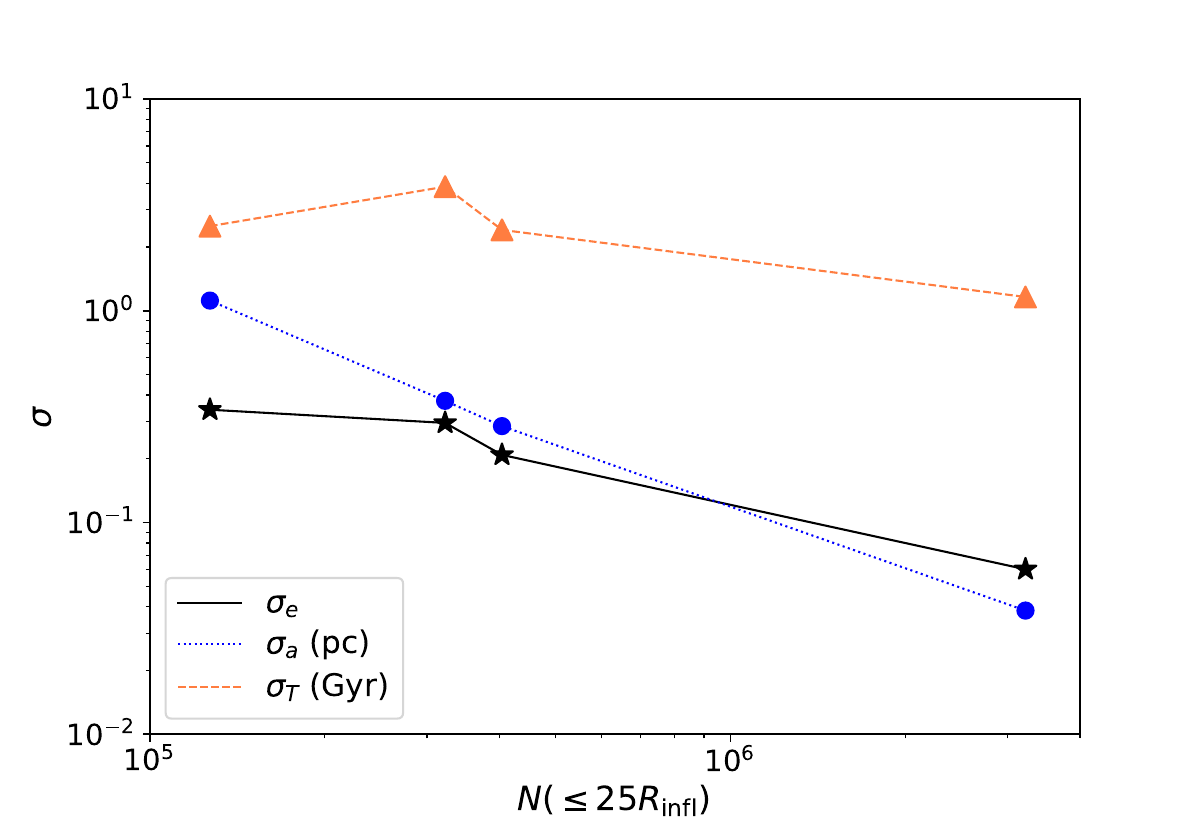}
	\caption{Dispersion of the BHB eccentricity (solid line), semi-major axis (dotted line) and merger timescale (dashed line) calculated from the random realisations of the $N=10^{6}$ models and the three M7 models, shown as a function of the number of particles within $25\bhsi$ of the binary (i.e for models M6, M6a, M6b, and M7s from left to right). The dispersion in all quantities decreases with increasing resolution, proving that the mass refinement schemes are effective. However, convergence in semi-major axis is significantly faster than in eccentricity, resulting in a large dispersion in merger timescales even at the highest resolution.}
     \label{fig:MNBH_dispersion}
\end{figure}

Table~\ref{tab:BHmergers} gives the time spent in each phase of evolution, namely dynamical friction, stellar hardening and GW emission. The latter two processes coexist in the hardening phase, we therefore define an arbitrary boundary between the two when the rate of change of semi-major axis due to each process is equal. All times are  calculated as total evolutionary times since the onset of the galactic merger. The lowest resolution models give timescales in the range $3-13\gyr$, with the longest being $\sim 13.3\gyr$ for model M6avi.  All high resolution models give timescales under $11\gyr$, a desirable result in terms of detection rates for PTA.

We calculate the dispersion in semi-major axis, eccentricity (both taken at time $T=2.6\gyr$) and merger timescales. These are shown in Fig.\,\ref{fig:MNBH_dispersion} for all models, as a function of the number of particles within $25\bhsi$ of the binary, as a measure of the effective resolution achieved in each model. We find that
the dispersion in all quantities decreases with increasing resolution (from models M6, to M6a, M6b and M7), meaning that the spread is reduced by the schemes and by increasing particle number. However, the dispersion in eccentricity is reduced much more slowly than the one in semi-major axis, due to its intrinsically stochastic nature. Because of the strong dependence of the merger timescale on initial binary eccentricity, convergence in the total timescale is not achieved even in models M7a/b, and would require even higher resolution and/or more random realisations to verify.

Finally, we have explored the effect of the choice of a fixed softening on convergence by re-running one realisation each of the M6, M6a, M6b models with a mass dependent softening $\epsilon$, according to $\epsilon_{bh} \propto m^{1/3}$. We find that the measured dispersion in eccentricity and merger timescale is largely unchanged.

\section{Discussion and Conclusions}
\label{sec:results}
We have presented a multi-mass refinement scheme to improve resolution at the centres of multi-component galaxy models, without impacting their stability and dynamical evolution. The initial model is over-seeded by a set amount, with both halo and bulge particles divided into radial zones. The defined central zone remains unchanged at the higher resolution produced by the over-seeding, whilst in the other radial zones we remove particles and correspondingly increase the masses of the remaining particles to compensate. In this way, the resulting galaxy model has a higher central resolution, necessary to accurately model MBH dynamics, at the same particle number, total stellar mass, and mass density profile.

We introduced two schemes with varying degrees of over-seeding, the standard model `a' and the more aggressive model `b', with `b' models having one more radial zone to compensate for the increased particle masses at outer radii. These implementations are applied to a set of isolated and merging multi-component galaxy models, each comprised of a dark matter halo, a stellar bulge, and a central MBH. We evolved a set of isolated and equal-mass merger models using the FMM code \textsc{griffin}, at the reference resolutions of $N=10^{6}$ and $N=10^{7}$, including a set of $7$ random realisations for each of the $N=10^{6}$ resolution models.

The isolated models show that the multi-mass schemes are effective at increasing resolution in the central regions, and this is maintained over time. Both schemes are also effective at reducing the relaxation-driven expansion of the bulge particles observed in the lower resolution models, and expansion is nearly eliminated in the M6b implementation. We show that the introduction of mass refinement does not affect the stability of the models nor the subsequent evolution. Density profiles, merger remnant shape and anisotropy profiles remain unchanged. Binary formation occurs at $\sim2.5\gyr$ for all models. However, while all models are able to reproduce the same behaviour for the separation between the MBHs over time, the orbital elements are sensitive to resolution.

The M6 models show a large scatter in semi-major axis compared to mass refined or higher resolution models. Models M6a and M6b are consistent with each other and with the M7 models.
The dispersion in semi-major axis calculated from the different random realisations shows convergence at a resolution of $N=10^6$, if a refinement scheme is applied.

The binary eccentricity, however, shows more variation than the semi-major axis, as it is more sensitive to perturbations and low-$N$ effects. A significant scatter is present at the lowest resolution of $N=10^6$ and some variation still exists at the highest resolution of $N=10^7$. Models M7a and M7b, however, have very similar eccentricity evolution.

We have computed the time to coalescence for the BHBs in all models by extending the $N$-body simulations with a semi-analytical calculation of the evolution of the orbital elements under the combined effects of stellar hardening and GW emission. We have determined the time-dependent hardening rate directly from the $N$-body integrations and used an exponential fit in the models. We find that merger timescales can vary by a several Gyrs in the M6 models and by a few Gyr in the M7 models. Model M6b, with the most aggressive scheme, gives the most similar evolution and merger timescale to the higher resolution models. By contrast, the $N=10^{7}$ models show a smaller spread in predictions, with models M7a and M7b coming within $\sim 1\gyr$ of each other. 

Variations in merger timescales have also been reported in \citet{Gualandris2022} at particle numbers of order $N=10^6$, while \citet{2021Nasim} show that a resolution in excess of $N=10^7$ within the half-light radius is required to reduce the scatter in the merger timescale to $\sim10\%$. We calculate that this corresponds to about $N=4\times10^6$ particles\footnote{all particles, whether stars or DM particles, are considered in this estimate, though bulge particles dominate at these small scales.} within $5\,\bhsi$, which is a factor 3 larger than what we achieve in models M7a/b. Taking the results of \citet{2021Nasim} at face value, this implies that the effective resolution in models M7a/b results in a $20-30\%$ error in the predicted merger timescale. Achieving a $10\%$ error is hence possible through a combination of $N(<5\bhsi)=4\times10^6$ and application of the mass refinement scheme.

We conclude that mass-refinement schemes of the type described in this work are extremely effective at increasing resolution in the central regions, and should be combined with an appropriate particle number to reduce scatter in key quantities: a particle number of $N(<5\bhsi)=6\times10^4$ is sufficient (e.g. model M6b) to reach convergence in the evolution of the semi-major axis, while a particle number of $N(<5\bhsi)=10^6$ is required to approach convergence in the evolution of the eccentricity and in the time to coalescence. In order to achieve a $10\%$ error on the merger timescale, a resolution $N(<5\bhsi)=4\times10^6$ is required. This is due to the strong dependence of both stellar hardening and GW emission on eccentricity. 

We note that our galaxies are all set on initial orbits of $e=0.7$, i.e. moderately eccentric. \citet{Gualandris2022} have shown that binaries with lower eccentricities exhibit a greater spread in merger timescales. We therefore expect somewhat lower variations for more eccentric orbits ($e\sim 0.9$) \citep[e.g.][]{Nasim2020, 2021Nasim,2022Mannerkoski, Rawlings2023}. 

Merger timescales can span a range from a few to several Gyr, making it likely that a third intruder will interact with the merger remnant before the BHB has reached coalescence. Subsequent mergers and triple galaxy encounters need to be modelled to make reliable predictions of merger timescales for PTA. A study of the merger tree selected from Illustris-TNG-300-1 presented in this work is underway. A refinement scheme of the type presented here is particularly beneficial for such studies given the need for high accuracy. While in this work we have only tested the mass refinement schemes on equal mass mergers of spherical galaxies, preliminary results show that they can be reliably applied to different morphologies (axisymmetric or triaxial models) and unequal mass mergers. This opens up the possibility of adopting mass refinement schemes as a best practice in any $N$-body simulation where large resolution is key, including cosmological simulations.

\section*{Acknowledgements}
We thank Tariq Hilmi and Matthew Orkney for interesting discussions and Eugene Vasiliev for support with the \textsc{agama} software.  The simulations were run on the {\tt Eureka} and {\tt Eureka2} HPC clusters at the University of Surrey.

\section*{Data Availability}
The data underlying this article will be shared on reasonable request to the corresponding author.



\bibliographystyle{mnras}
\bibliography{BIBLIO} 




\appendix


\bsp	
\label{lastpage}
\end{document}